\documentclass[a4paper,amsfonts,amssymb,amsmath,reprint,floatfix,nofootinbib,twoside,superscriptaddress]{revtex4-2}
\usepackage[english]{babel}
\usepackage[utf8]{inputenc}
\usepackage{amsthm}
\usepackage{mathtools}
\usepackage{physics}
\usepackage{xcolor}
\usepackage{graphicx}
\usepackage[left=23mm,right=13mm,top=35mm,columnsep=15pt]{geometry} 
\usepackage{adjustbox}
\usepackage{placeins}
\usepackage[T1]{fontenc}
\usepackage{lipsum}
\usepackage{csquotes}
\usepackage[pdftex,hidelinks,pdftitle={Why shaking faster does not untangle},pdfauthor={Ishant Tiwari, Sutikshan Bansal, Rosalind Huang, Vishal Patil, Chenyi Fei, Jorn Dunkel, and Saad Bhamla}]{hyperref}
\usepackage{xcolor}

\begin{document}

\title{Why shaking faster does not untangle:\\ frequency-selective extraction of filaments from tangled matter}
\author{Ishant Tiwari}
\email[Correspondence: ]{ishant@coloradocollege.edu}
\affiliation{Department of Physics, Colorado College, Colorado Springs, Colorado 80903, USA}
\affiliation{School of Chemical and Biomolecular Engineering, Georgia Institute of Technology, Atlanta, Georgia 30332, USA}

\author{Sutikshan Bansal}
\thanks{These authors contributed equally.}
\affiliation{School of Materials Science and Engineering, Georgia Institute of Technology, Atlanta, Georgia 30332, USA}

\author{Rosalind Huang}
\thanks{These authors contributed equally.}
\affiliation{Department of Physics, Massachusetts Institute of Technology, Cambridge, Massachusetts 02139, USA}

\author{Chenyi Fei}
\affiliation{Department of Mathematics, Massachusetts Institute of Technology, Cambridge, Massachusetts 02139, USA}

\author{J\"orn Dunkel}
\affiliation{Department of Mathematics, Massachusetts Institute of Technology, Cambridge, Massachusetts 02139, USA}

\author{Vishal Patil}
\affiliation{Department of Mathematics, University of California San Diego, La Jolla, California 92093, USA}

\author{Saad Bhamla}
\email[Correspondence: ]{saad.bhamla@colorado.edu}
\affiliation{BioFrontiers Institute and Department of Chemical and Biological Engineering, University of Colorado Boulder, Boulder, Colorado 80309, USA}
\date{October 13, 2026}

\begin{abstract}
Tangles of soft filaments are ubiquitous, from living worm blobs to the cables in a desk drawer. How do such tangles, held together by friction and entanglement, respond when one embedded filament is shaken? We suspend disordered tangles of tunable cohesion and ordered stitched meshes from a single thread (the tether) and oscillate it vertically at fixed amplitude. Extraction peaks at an intermediate drive frequency ($\sim15$~Hz) and declines at higher frequencies, although the peak drive acceleration grows quadratically. Here we combine experiments, high-speed imaging, and a tension-only spring--mass model to show that this non-monotonic response is set by timing rather than resonance. The extraction window is bounded below by a slackening threshold, where the peak drive acceleration exceeds gravity and the tether, slack for part of each cycle, creates a tensile impulse each time it snaps taut. The window is bounded above by a transmission crossover, where the transmission time across the tangle exceeds a quarter of the drive cycle, so that the tangle no longer responds as a whole and the drive’s work dissipates internally. The slackening threshold depends only on the drive and gravity, while the transmission crossover is set only by the tangle’s size and transmission speed. By tracing extraction to the overlap of impulse creation and impulse transmission, we identify a generic principle of driven filamentous matter and suggest frequency-selective routes to its disassembly.

\end{abstract}

\maketitle

\section*{Introduction}

Entanglement determines the cohesion and mechanics of filamentous matter across scales, from bird nests of twigs~\cite{weiner2020mechanics,bhosale2022micromechanical} to entangled metal rubbers~\cite{zhang2014dynamic} and neurofibrillary tangles~\cite{itoh1997scanning,rankin2007tau}. In these disordered tangled materials,  macroscopic rigidity arises from  geometry, topology, contact constraints, and friction acting together, rather than from the elasticity of the individual filaments~\cite{jung2025entanglement,patil2020topological}. The resulting networks can be cohesive without adhesives, plastic despite elastic constituents, and soft even though every filament is rigid.

Agitation continually reorganizes these networks. For example, shaken strings and chains knot themselves spontaneously~\cite{raymer2007spontaneous,belmonte2001dynamic,soh2019self,gendron2021time}, vibration unties knots in granular chains~\cite{bennaim2001knots}, and the same shaking that collapses a weakly entangled rod packing consolidates a strongly entangled one~\cite{jung2025entanglement}. Even a single tight contact or a physical knot derives its holding force from the interplay of elasticity and friction~\cite{grandgeorge2021orthogonal,johanns2023surgical,johanns2024capsizing}. 
Living tangles exploit this reorganization, with blobs of slender worms  agitating themselves to untangle within tens of milliseconds~\cite{patil2023ultrafast}. For amorphous solids, low-frequency vibrational modes  connect structural disorder to the driven response~\cite{hu2022origin,wyart2005effects,chen2010low}. To our knowledge, no comparable description exists for an amorphous tangle,  whose elementary motions  extend beyond small-amplitude modes to finite rearrangements. Bending, sliding, contact switching, and loop opening allow a localized oscillation either to reorganize the contact configuration or to dissipate as internal deformation. 

 Anyone who has struggled to shake a single cable free from a tangled collection of wires in a desk drawer has applied a localized oscillation to a topologically entangled  network of frictional  contacts. A slender filament supports tensile load yet buckles under the slightest compression, so a driven cable  alternates between pulling on the tangle when taut and letting it fly freely when slack. Free flight begins once the peak drive acceleration exceeds gravity, as it does for a ball on a vibrating plate or a vibrated granular layer~\cite{melo1995hexagons,tufillaro1986bouncing}. This unilateral loading also places the driven tangle among bilinear and impact oscillators, which convert smooth harmonic forcing into nonsmooth, strongly nonlinear dynamics~\cite{shaw1983periodically,thompson1983subharmonic,nordmark1991grazing,ing2008impact}. A tangle, however, is an extended medium as well as an oscillator, and a tensile impulse created when the tether snaps taut must travel outward through a network of contacts that can rearrange~\cite{jia1999ultrasound,ma2023experimental}.

When a tangle  hangs from a single  filament embedded within it and that filament is shaken, what governs its motion? And is there an optimal way to shake the filament free? We address these questions with disordered thread tangles whose cohesion we tune by tumbling them in a ``tangler''. An ``extractor'' vertically oscillates one randomly chosen filament, the tether, at a tunable frequency. High-speed imaging then tracks how a disturbance initiated at the tether spreads into the surrounding tangle. Combining these observations with a tension-only spring--mass model, we show that extraction is confined to an intermediate forcing window between two boundaries: a slackening threshold, below which the tether never falls slack and no tensile impulse is created, and a transmission crossover, where the  travel time of an  impulse becomes comparable to the drive period. Tracing the first boundary to impulse creation and the second to impulse transmission, we identify a generic dynamical principle of driven filamentous matter that can guide the selective disassembly of tangles by shaking.

\section*{Results}

\begin{figure*}
    \centering
    \includegraphics[width=\linewidth]{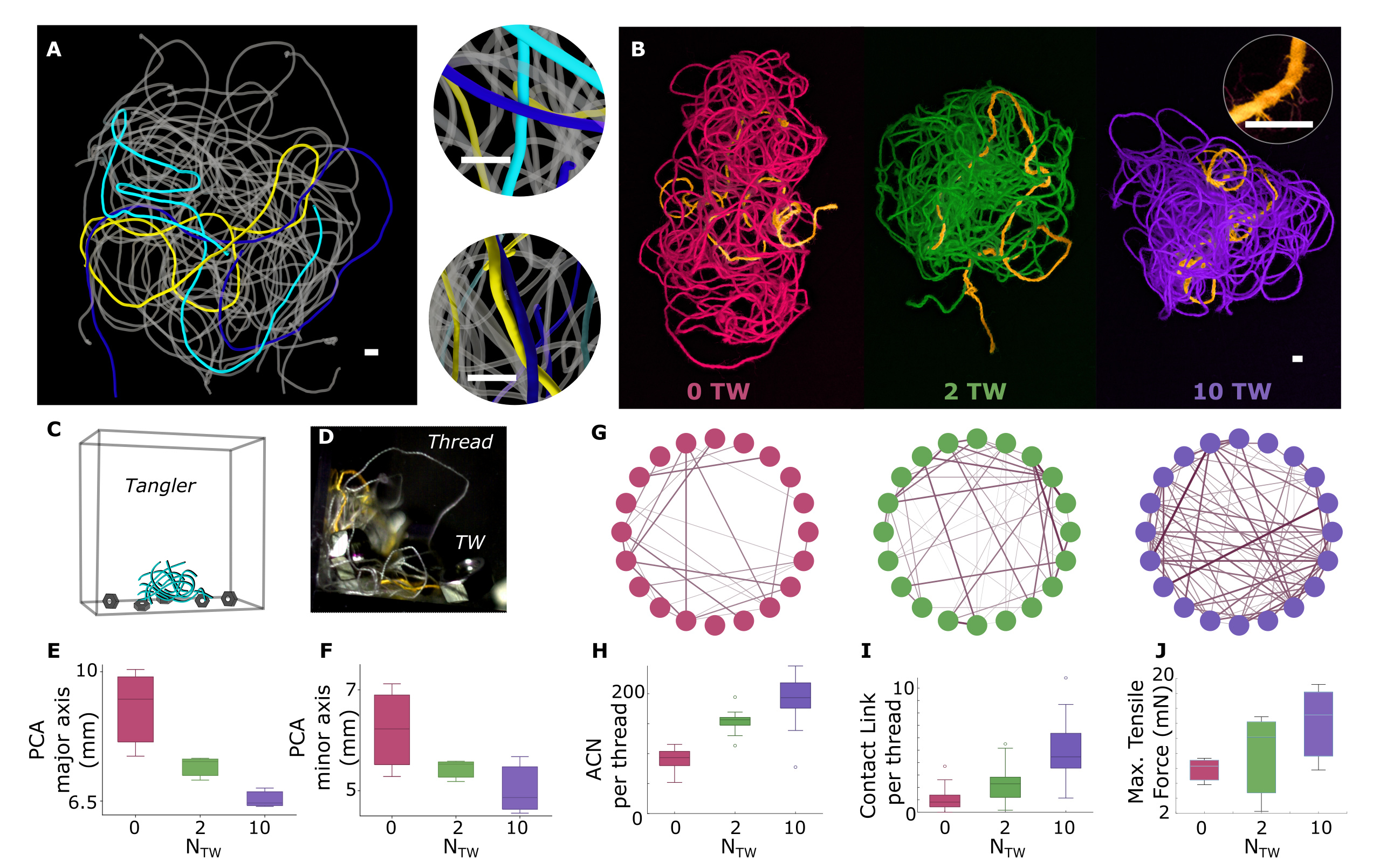}
    \caption{\textbf{The tangler produces cohesive topological tangles with tunable entanglement.}
    \textbf{A} Micro-CT reconstruction of a representative cotton-thread tangle, with enlarged views of transverse crossings, and longer tangentially aligned contacts.
    \textbf{B} Representative tangles produced with $N_{TW}=$ 0, 2 and 10 tumbling weights (TW).
    \textbf{C--D} The tangler, a rotating cuboidal box, with a representative
cotton thread and tumbling weight;  each tangle contains 20 threads of length $10\,$cm. 
\textbf{E--F} Major- and minor-principal-axis root-mean-square (RMS) extents of the projected tangle pixels ($n=3$ independently prepared tangles per TW condition).
\textbf{G} Contact graphs extracted from one micro-CT reconstruction per condition; nodes are threads and edges denote identified contacts, with thickness and opacity scaled by the square root of the magnitude of the open Gauss linking between contacting pairs~\cite{patil2023ultrafast}, normalized across the three conditions.
    \textbf{H--I} Average crossing number (ACN)~\cite{becker2022active} per thread and contact linking number (linking between touching threads~\cite{patil2023ultrafast}) per thread as a function of $N_{\rm TW}$ ($n=1$ micro-CT scan per TW condition; box plots denote thread level values within each reconstructed tangle).
    \textbf{J} Peak force in quasi-static single-thread pullout tests ($n=3$ independently prepared tangles per condition). For \textbf{E--F and H--J}, the center line denotes the median, the box spans the interquartile range, the whiskers extend to
the most extreme non-outlier values, while individual points are the outliers. Scale bars, 1mm }
    \label{fig:tangle_creation}
\end{figure*}

\subsection*{Tumbling weights tune tangle topology and cohesion}

A first challenge in studying optimal extraction is producing statistically repeatable tangles from individual filaments (Fig.~\ref{fig:tangle_creation}A,B). We produced them with a ``tangler,'' a rotating cuboidal box that agitates 20 threads of length $10\,\mathrm{cm}$ together with a prescribed number $N_{\rm TW}$ of tumbling weights (TW; Fig.~\ref{fig:tangle_creation}C,D and Supplementary Video~1). This preparation extends the rotating-container protocol that first produced spontaneous  knotting of a single string~\cite{raymer2007spontaneous}, with the added weights tuning how strongly the threads are driven into one another and how readily they link. Micro-CT reconstructions reveal transverse crossings as well as longer, tangentially aligned contacts (Fig.~\ref{fig:tangle_creation}A). These motifs provide the normal constraints and extended frictional interfaces through which a force applied to one thread spreads across the aggregate.

Across the three preparation conditions, increasing $N_{\rm TW}$ from 0 to 2 to 10 compacts the tangle even as the thread count remains fixed at 20 (Fig.~\ref{fig:tangle_creation}B). We quantify this compaction from two-dimensional projections by measuring their major and minor-principal-axis RMS extents (Fig.~\ref{fig:tangle_creation}E-F; Supplementary Information, section III). The micro-CT reconstructions give the same ordering in three dimensions, with envelope diameters of 28.6, 24.4 and 23.7\,mm for 0, 2 and 10 TW.

The contact architecture reorganizes along with this compaction. Reducing each  reconstructed tangle to a graph, in which a node represents one thread and an edge records at least one contact between a pair of threads, shows the network becoming  more connected with more tumbling weights (Fig.~\ref{fig:tangle_creation}G). The topological entanglement also grows, with both the average  crossing number (ACN)~\cite{becker2022active} per thread and the contact linking number~\cite{patil2023ultrafast} per thread shifting upward with stronger tumbling (Fig.~\ref{fig:tangle_creation}H--I).

This growing entanglement has a direct mechanical consequence. When we pull  one embedded thread quasi-statically and record the largest tensile force before release,  the median peak pulling force increases with $N_{\rm TW}$ (Fig.~\ref{fig:tangle_creation}J; $n=3$ independently prepared tangles per condition). Tumbling thus supplies a reproducible progression from a loose aggregate to a compact, topologically complex tangle capable of carrying a localized tensile load. Having characterized the tangles at rest, we now ask  how they respond when we drive a single embedded thread dynamically.

\begin{figure*}
    \centering
    \includegraphics[width=\linewidth]{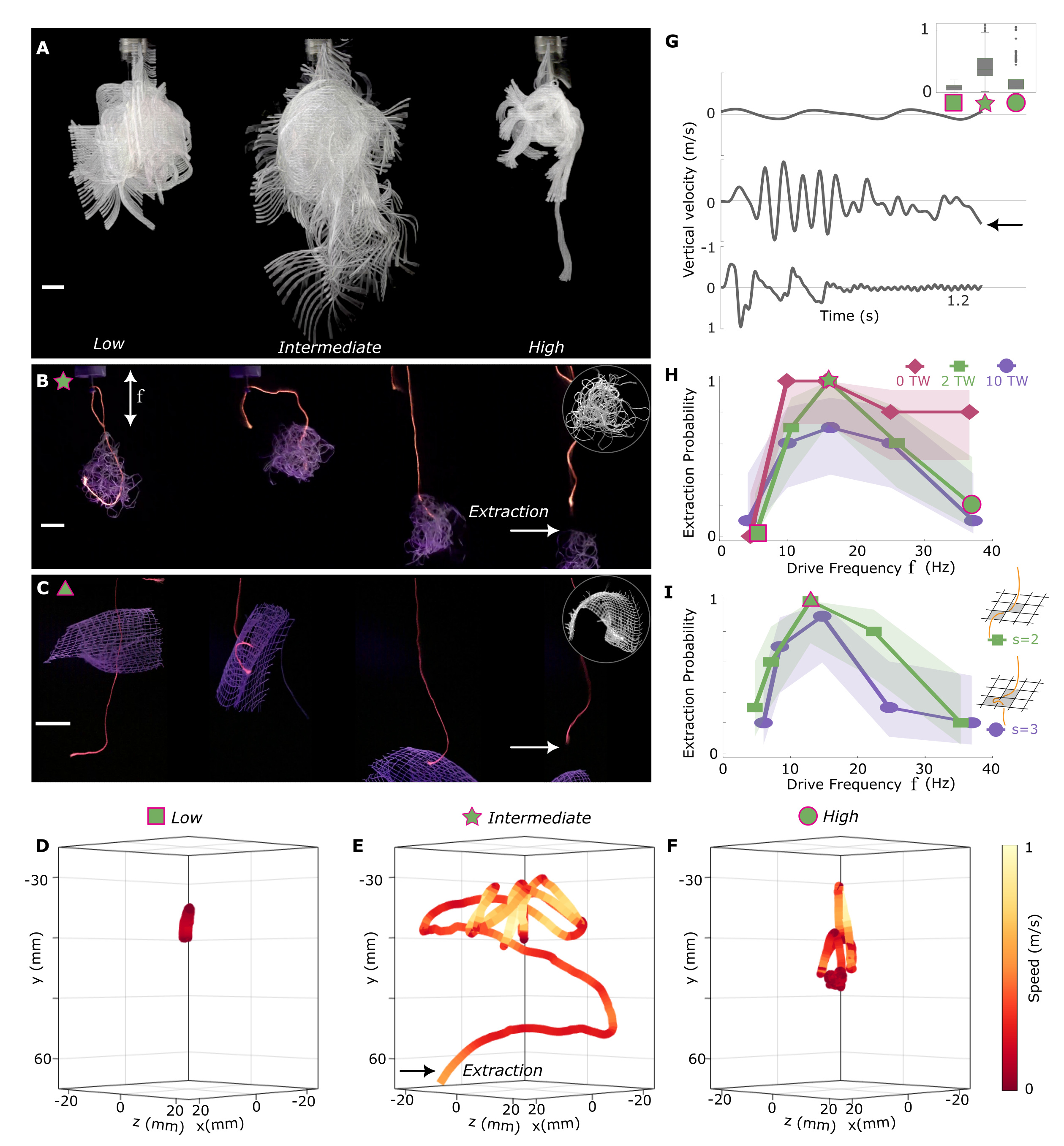}
    \caption{\textbf{The extractor reveals three driven regimes and an  intermediate extraction window.}
    \textbf{A} Overlaid frames of a disordered tangle at representative low, intermediate, and high frequencies.
    \textbf{B} Representative sequence in which repeated slip culminates in extraction from a disordered tangle (full dynamics  in Supplementary Video~2). The vertical double-headed arrow denotes the drive direction, whereas the horizontal arrow marks extraction. 
    \textbf{C} Corresponding sequence for a ordered stitched mesh; the horizontal arrow again marks extraction.
    \textbf{D--F} Three-dimensional center-of-mass trajectories for the low-, intermediate- and high-frequency regimes, colored by instantaneous speed; the horizontal arrow in \textbf{E} marks extraction.
    \textbf{G} Vertical center-of-mass velocity for the three representative trials, with vertically offset zero lines; the arrow marks extraction. Inset: distribution over each trial of the absolute vertical center-of-mass velocity $|v_y|$ (m\,s$^{-1}$) for the three regimes (center line, median; box, interquartile range; whiskers, most extreme non-outlier values).
    \textbf{H} Probability $P_{\rm ext}$ of complete extraction within $300\,$s among $n=10$ independent trials for 0-, 2- and 10-TW tangles.
    \textbf{I} $P_{\rm ext}$ for ordered meshes with stitch count $s=2$ or $s=3$; the schematics show the two stitching paths. Each probability contains $n=10$ independent trials. The shaded bands in H--I are two-sided 95\% Wilson binomial confidence intervals on the observed extraction fraction. Scale bars, 1~cm. Throughout, square, star, and circle markers identify the representative low, intermediate, and high frequencies of \textbf{A}--\textbf{G} on the probability curves. The triangle identifies the drive frequency of the representative stitched-mesh sequence in panel C.}
    \label{fig:untangling}
\end{figure*}

\subsection*{Filament extraction is maximal in an intermediate forcing window}

We suspend each tangle from a randomly chosen embedded thread (the tether) and drive the suspension point  with the extractor, a piston sliding in a vertical sleeve, at fixed amplitude $A=0.75\,\mathrm{cm}$ (Fig.~\ref{fig:untangling}A). Scanning over the forcing frequency produces three qualitatively distinct responses. At low frequency, the aggregate follows the piston in a narrow vertical bob. At intermediate frequency, $f\simeq13$--$16\,\mathrm{Hz}$, the tangle makes large irregular excursions punctuated by abrupt tensioning and rapid motion. These excursions culminate in the tether pulling free entirely and the tangle slipping free (Fig.~\ref{fig:untangling}B and Supplementary Video~2). At high frequency, $f\simeq35$--$40\,\mathrm{Hz}$, the tangle's center of mass localizes again even though individual regions continue to deform. The global response is therefore non-monotonic even as the drive acceleration rises monotonically with frequency. 

Three-dimensional center-of-mass trajectories quantify this contrast (Fig.~\ref{fig:untangling}D--F). While the low-frequency trajectory  again follows the piston's motion closely, the intermediate-frequency trajectory explores the largest spatial volume and contains the fastest segments, and the high-frequency trajectory collapses into a compact lobe after an initial startup jolt. In the vertical center-of-mass velocity (Fig.~\ref{fig:untangling}G), only the intermediate-frequency trial sustains large fluctuations until extraction. Large global excursions  do not by themselves free the thread. They do, however, repeatedly load and unload the tether, and thereby create opportunities for slip.

We measure the outcome  through $P_{\rm ext}$, the fraction of $n=10$ trials in which the tether is completely extracted within $300\,\mathrm{s}$. The 2- and 10-TW tangles show an intermediate-frequency maximum, with extraction becoming less likely on either side (Fig.~\ref{fig:untangling}H). By contrast, the weakly cohesive 0-TW tangles remain extractable over a broader range, consistent with their smaller quasi-static pullout forces (Fig.~\ref{fig:tangle_creation}J). An extraction window therefore emerges once the aggregate can sustain an appreciable load.

To test how this window depends on contact-network structure, we replaced the disordered tangle with an ordered gauze sheet and stitched the tether through it along prescribed paths (Fig.~\ref{fig:untangling}C). Meshes with $s=2$ and $s=3$ stitches recover an intermediate maximum in $P_{\rm ext}$ (Fig.~\ref{fig:untangling}I), even though they  lack the random packing and  distributed contact network of the disordered tangles. Although the measured stiffness-to-weight ratio $\chi=k_{\rm eff}L_0/(mg)$ varies fourfold across the tangles and meshes ($\chi=184$--$755$; $k_{\rm eff}$ is the quasi-static loading slope; Supplementary Information, sections IV and VI)), the extraction window remains at $f\simeq13$--$16\,\mathrm{Hz}$, significantly below the lumped natural frequencies calculated from every measured quasi-static stiffness (ex. $\approx50\,\mathrm{Hz}$ for the 10-TW tangle). The relevant physics must therefore be common both to the tangle and the mesh. Each hangs an extended frictional body from a compliant filament that can pull but cannot push.   In both,  the drive must  create a tensile impulse at the tether, and the suspended body must transmit it. We examine these two requirements in turn.

\subsection*{Slackening and transmission bound the extraction window}

\begin{figure*}[t]
    \centering
    \includegraphics[width=0.94\linewidth]{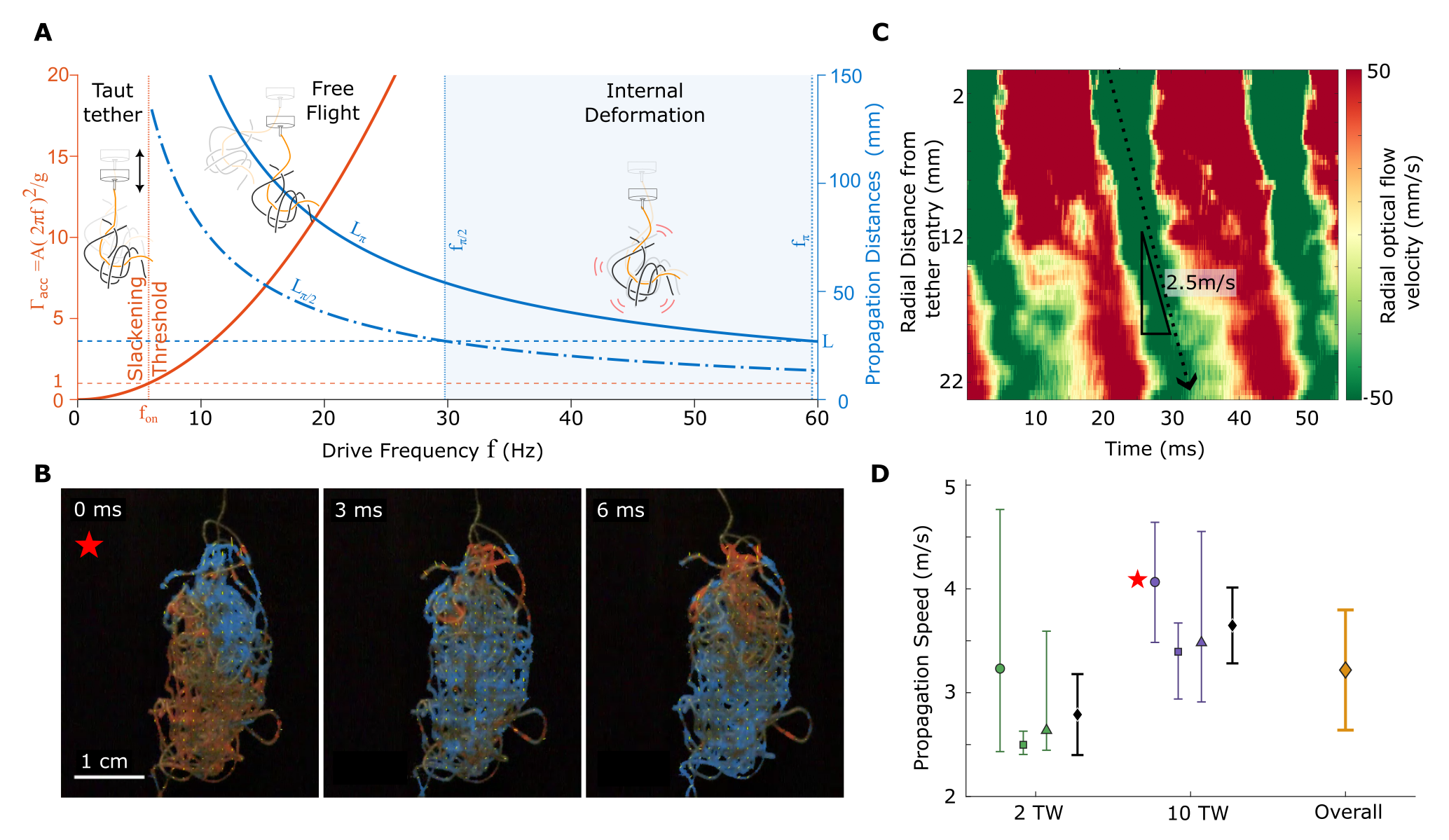}
    \caption{\textbf{The slackening threshold and the transmission crossover bound the extraction window.}
    \textbf{A} Peak drive acceleration $\Gamma_{\rm acc}=A(2\pi f)^2/g$ (orange, left axis) and distances traveled in a quarter cycle, $L_{\pi/2}=v_{\rm eff}/(4f)$ (blue dash--dotted), and half cycle, $L_\pi=v_{\rm eff}/(2f)$ (blue solid, right axis), at $A=0.75\,$cm. The orange guides mark $\Gamma_{\rm acc}=1$ and $f_{\rm on}=5.8\,$Hz. The horizontal blue guide is the pooled projected extent $L_{95}=27\,$mm. Its intersections with the two timing curves give $f_{\pi/2}=29.8\,$Hz and $f_\pi=59.5\,$Hz, which bound the shaded transmission crossover window. Sketches indicate the taut-tether, free-flight, and internal-deformation regimes.
    \textbf{B} High-speed optical-flow snapshots showing motion spreading from the tether entry point toward the tangle periphery (10 TW tangle, complete sequence in Supplementary Video~3). Cyan and orange shading indicate the measured radially inward and radially outward local optical-flow field respectively. The red star identifies this recording in D. Scale bar, 1~cm.
    \textbf{C} Signed radial-velocity kymograph from a high frequency recording with a 2 TW tangle; the slopes of the oblique ridges of radially inward motion (green) measure the outward progression of deformation. The dotted guide illustrates a propagation speed of 2.5~m/s.
    \textbf{D} Hierarchical propagation speed summary measured with n=3 repeats each for 2 TW and 10 TW conditions. Colored circles, squares and triangles denote R1, R2 and R3 recording medians, with thin whiskers spanning their event-level interquartile ranges. Black diamonds show condition means with $\pm1$ sample standard deviation across three recording medians. The gold diamond is the overall mean, $3.2\pm0.6\,{\rm m\,s^{-1}}$, with the standard deviation across all six recording medians.}
    \label{fig:regimes}
\end{figure*}

As  the drive amplitude is fixed in our experiments, scanning the forcing frequency also scans the peak acceleration of the support, which grows as $f^2$. An ideal tether  slackens whenever its tension  reaches zero, and for a suspension point moving as $y_s=A\sin(2\pi f t)$, the minimum tension $T_{\rm min}$ over a drive cycle follows,
\begin{equation}
T_{\min}=\max\{0,\,mg\left(1-\Gamma_{\rm acc}\right)\},\qquad
\Gamma_{\rm acc}=\frac{A(2\pi f)^2}{g},
\label{eq:slack_threshold}
\end{equation}
where $\Gamma_{\rm acc}$ is the dimensionless peak drive acceleration
familiar from vibrated granular media~\cite{melo1995hexagons}. For $\Gamma_{\rm acc}<1$, gravity keeps the tether taut and the tangle aggregate follows the piston without episodes of free flight (Supplementary Video 2). The tension first becomes zero  at $\Gamma_{\rm acc}=1$, which at our amplitude of $0.75\,$cm  occurs at $f_{\rm on}=5.8\,\mathrm{Hz}$, close to the observed onset of extraction (Fig.~\ref{fig:regimes}A, vertical orange dashed line). We call $\Gamma_{\rm acc}=1$ the slackening threshold. Above it, the tether goes slack each time  the piston moves toward the tangle; the released tangle travels freely until the tether snaps taut again, converting the smooth drive into intermittent tensile impulses. Crossing the slackening threshold therefore satisfies the first requirement for extraction. Whether an impulse then succeeds in extraction depends on frictional slip between filaments and how the extended tangle transmits force through its network.

The peak acceleration $\Gamma_{\rm acc}$ by itself cannot explain the change in the tangle's response at high drive frequency. Indeed, $\Gamma_{\rm acc}$ continues to grow as $f^2$, while both the center-of-mass motion and the extraction probability fall. To follow how perturbations propagate through the tangle, we record at $10{,}000$ frames per second and project the optical-flow field radially from the
tether entry point (n=3 independent trials each at 2 TW and 10 TW; Supplementary Information, section V). At each recapture, motion begins near the entry point and spreads toward the periphery over several milliseconds (Fig.~\ref{fig:regimes}B; Supplementary Video~3). The resulting kymograph shows oblique bands whose slopes measure the outward progression of the impulse (Fig.~\ref{fig:regimes}C). From the slopes of these bands we obtain $v_{\rm eff}=2.8\pm0.4\,{\rm m\,s^{-1}}$ for the 2-TW and $3.6\pm0.4\,{\rm m\,s^{-1}}$ for the 10-TW tangles (mean $\pm$ s.d. of three recording medians per condition; Supplementary Information, section V), and $v_{\rm eff}=3.2\pm0.6\,{\rm m\,s^{-1}}$ over all six recordings (Fig.~\ref{fig:regimes}D).
 This speed describes how deformation travels  through a disordered contact network and sets the transit time across the suspended tangle aggregate, much as measured transmission speeds do in stressed granular
packings~\cite{jia1999ultrasound} and entangled metallic wires~\cite{ma2023experimental}.

To compare this transit time with the forcing, we formulate the accumulated phase 
\begin{equation}
\phi=\frac{2\pi f L}{v_{\rm eff}},
\label{eq:propagation_phase}
\end{equation}
which measures the delay across a tangle of extent $L$ in radians. We take $L=2R_{95}$, where $R_{95}$ is the radius that encloses 95\% of a tangle’s projected area about its centroid. Across three 2-TW and three 10-TW tangles, $L=27\pm1.8\,$mm (mean $\pm$ sample standard deviation; Supplementary Information, sections III and V). 

For $\phi\ll\pi$, recapture can move the extended body approximately coherently (the perturbation spans the tangle within a small fraction of the cycle). We define the transmission crossover window using quarter- and half-cycle delays, $\phi=\pi/2$ and $\phi=\pi$:
\begin{equation}
    f_{\pi/2}=\frac{v_{\rm eff}}{4L}\simeq29.8\,{\rm Hz},
    \qquad
    f_\pi=\frac{v_{\rm eff}}{2L}\simeq59.5\,{\rm Hz}.
    \label{eq:timing_window}
\end{equation}
 The high-frequency trials,  $f\simeq35$--$40\,\mathrm{Hz}$, fall within this interval, and internal deformation, rather than coherent center-of-mass motion, increasingly dominates their dynamics (Fig.~\ref{fig:regimes}A). The tangle's far side is then like the bottom of a released slinky, which hangs motionless until the collapse front arrives from above~\cite{cross2012slinky}.

Since coherent response requires $\phi \ll \pi$, lowering  $\phi$ should preserve the high-frequency response.  As $\phi\propto L/v_{\rm eff}$, we can lower $\phi$ in two ways, by shortening the transmission distance $L$ or by raising the transmission speed $v_{\rm eff}$. Compacting the same tangle into a tight, mass-matched body shortens $L$, and  the compacted tangle indeed remains agitated  at high frequency (Supplementary Video~4). Replacing the compliant tangle of filaments with a mass-matched rigid plastic mesh, with a  filament stitched through it, raises $v_{\rm eff}$,  and the rigid mesh likewise remains agitated at high frequency  (Supplementary Video~5).
Conversely, increasing $L$ should move the crossover to lower frequency. A stitched mesh with a long ``wingspan'', driven at an intermediate frequency, deforms internally and does not move as a whole; folding the wings to shorten $L$ at the same mass restores the slack--recapture motion of the center of mass (Supplementary Video~6).
These controls behave as the transmission crossover predicts: with a
shorter transit time, a larger fraction of the body responds before the drive reverses and vice versa. Whether the drive creates a tensile impulse at all depends on its acceleration and the tangle's stiffness, a question we address next with a minimal mathematical model.

\subsection*{A minimal model separates impulse creation from impulse transmission}

\begin{figure*}
    \centering
    \includegraphics[width=\linewidth]{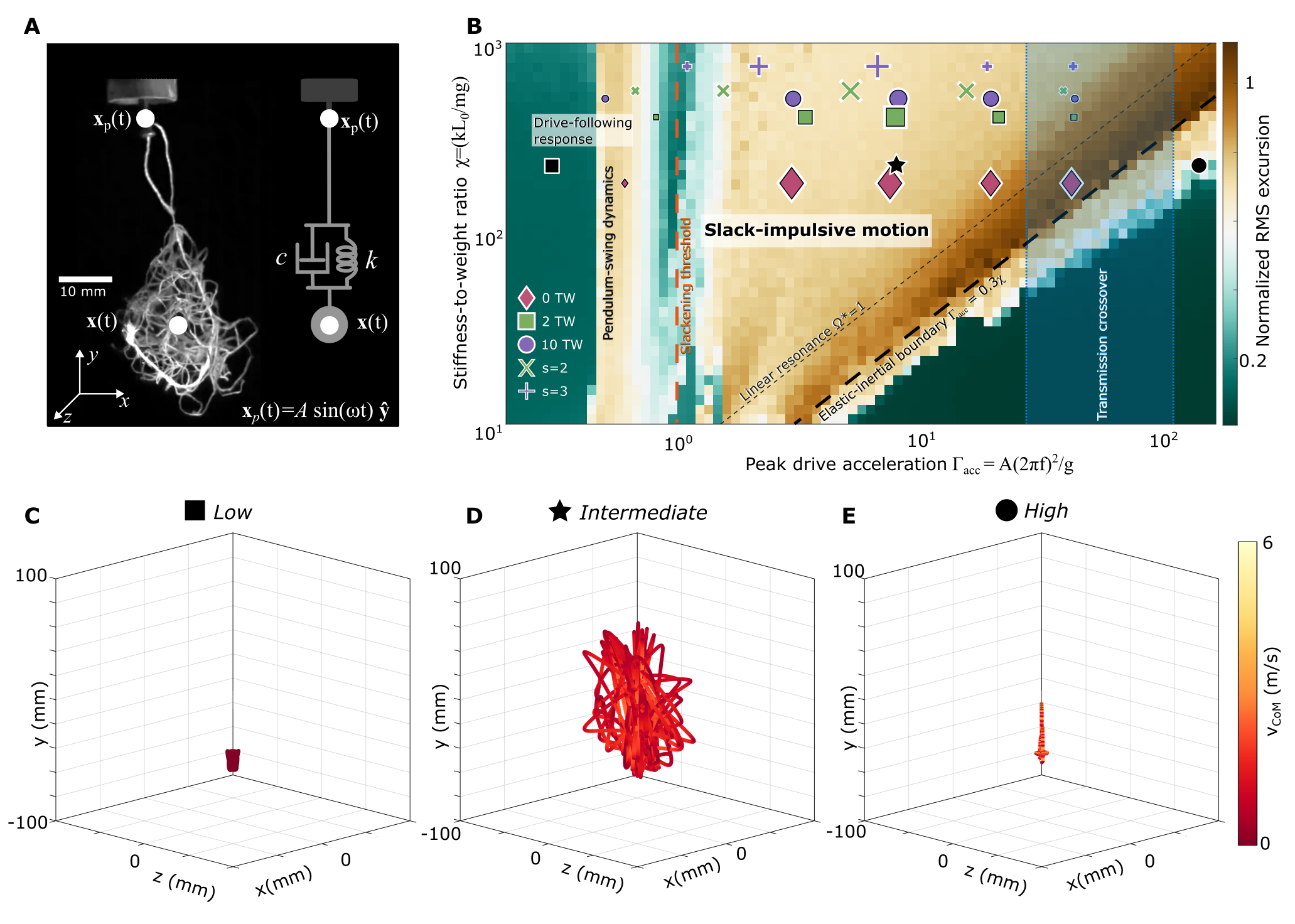}
    \caption{\textbf{The measured extraction probability peaks where the minimal model predicts slack–impulsive motion.}
    \textbf{A} Suspended tangle and its tension-only reduced-order representation: a mass at $\mathbf{x}(t)$ hangs from the driven support $\mathbf{x}_p(t)=A\sin(\omega t)\,\hat{\mathbf{y}}$ through a spring of stiffness $k$ active only in tension, a damper $c$, and a small fluctuating force (Supplementary Information section VI).
    \textbf{B} Ensemble-averaged  root mean square excursion $\langle A_{\rm rms}\rangle/L_0$ (Supplementary Information section VI) from a direct $80\times40$ scan of the $(\Gamma_{\rm acc},\chi)$ plane, with  ten noise realizations per point. The orange vertical dashed line marks the slackening threshold $\Gamma_{\rm acc}=1$; the thick dashed diagonal line is the lumped elastic--inertial boundary $\Gamma_{\rm acc}=2\epsilon\chi=0.30\chi$; the
thin dashed diagonal is the linear-spring resonance $\Omega^*=1$; and the two vertical blue boundaries delimit the quarter-to-half-cycle transmission crossover window of Fig.~\ref{fig:regimes}A, mapped to $\Gamma_{\rm acc}\simeq26.8$ and $107.1$. Experimental conditions (legend) are placed by their quasi-static stiffness-to-weight
ratios, with marker area proportional to the measured $P_{\rm ext}$. The black square, star, and circle mark the parameters of panels \textbf{C}--\textbf{E}.
    \textbf{C--E} Representative simulated trajectories in the low-, intermediate- and high-frequency regimes, colored by center-of-mass speed relative to the moving support (Supplementary Video~7). Panels \textbf{C--E} trajectories are plotted at the same scale.}
    \label{fig:model}
\end{figure*}

To decouple the effects of tangle stiffness and drive acceleration, we represent the suspended tangle as an effective mass joined to the moving support by a damped tension-only element (Kelvin-Voigt-like damping), in essence a yo-yo shaken from above (Fig.~\ref{fig:model}A). If $r$ is the instantaneous separation, the elastic contribution takes the form
\begin{equation}
T=k\max(0,r-L_0),
\label{eq:tension_only}
\end{equation}
so that the connector loads the mass in tension once stretched beyond its rest length $L_0$, but exerts no compressive force after slackening. The remaining  terms account for gravity, isotropic drag, radial damping, and small fluctuations, with their implementation  given in the Supplementary Information (SI section VI). This abstraction isolates the slack--recapture response and center-of-mass motion without prescribing any particular contact network configuration.

The model separates the drive from the tangle through two independent controls. The first is the peak drive acceleration $\Gamma_{\rm acc}$ of Eq.~\eqref{eq:slack_threshold}. The second is the stiffness-to-weight ratio
\begin{equation}
\chi=\frac{kL_0}{mg},
\label{eq:independent_coordinates}
\end{equation}
which compares the spring force at an extension $L_0$ with the suspended weight and is therefore dimensionless. In the response map, the reference length is $L_0=5\,$cm (half the length of an individual thread) and the amplitude is fixed at the experimental value $A=0.75\,$cm, so $\epsilon=A/L_0=0.15$. With $\epsilon=A/L_0$ and the elastic--inertial frequency ratio $\Omega^*=\omega\sqrt{m/k}$, the  two are linked by the exact relation
$\Gamma_{\rm acc}=\epsilon\chi(\Omega^*)^2$,
so lines of constant normalized frequency $\Omega^*$ run diagonally across the $(\Gamma_{\rm acc},\chi)$ plane.

We quantify the model's response at each point of this plane through the root-mean-square (RMS) excursion of the hanging mass about its temporal mean, $\langle A_{\rm rms}\rangle/L_0$ (Supplementary Information section VI). Small values correspond to localized motion, whereas large values indicate large RMS excursions generated by repeated slackening and recapture. Representative trajectories show how this observable distinguishes the three dynamical regimes (Fig.~\ref{fig:model}C--E).

Scanned over the $(\Gamma_{\rm acc},\chi)$ plane, this observable produces the response map of Fig.~\ref{fig:model}B. The slackening threshold $\Gamma_{\rm acc}=1$ appears as its vertical boundary, below which the simulated spring stays in tension and the mass largely follows the imposed  motion of the support, although at certain drive frequencies  it exhibits a pendulum-like response. Beyond the threshold,  alternating slack and taut phases generate large RMS excursions and repeated recapture events, corresponding to the slack--impulsive motion that accompanies successful extraction in the experiments. At still larger $\Gamma_{\rm acc}$, the response falls  along a diagonal boundary. This edge approximately follows a constant $\Gamma_{\rm acc}/\chi$, and hence, since $\Gamma_{\rm acc}=\epsilon\chi(\Omega^*)^2$, a constant elastic--inertial frequency ratio $\Omega^*$.

To estimate this frequency, we linearize the model during a taut interval about its gravity-stretched equilibrium. As long as the string stays taut, the pre-tension supplied by the hanging weight makes the one-way connector act as an ordinary two-sided spring, and the yo-yo  reduces to a point mass hanging from a shaken spring. The fraction of the support's motion transmitted to the mass then follows the classical transmissibility of base excitation. At low frequency the mass follows the support, and near resonance it overshoots. Driven rapidly, however, it hangs nearly still while the support oscillates above it. For the weak damping used here,  the mass's motion matches  the support's own amplitude at $\Omega_c^*\simeq\sqrt{2}$ (Supplementary Information section VIII). The identity $\Gamma_{\rm acc}=\epsilon\chi(\Omega^*)^2$ then places the boundary at $\Gamma_{\rm acc,el}\simeq2\epsilon\chi=0.30\chi$, the diagonal elastic--inertial guide in Fig.~\ref{fig:model}B.  For comparison, we also draw the corresponding diagonal for a linear spring, whose resonance lies at $\Omega^*=1$. Supplementary Video~7 shows representative model responses and their slack--recapture dynamics.

Finally, overlaying the experimental data on this map tests how much our point-mass model can explain. The largest measured extraction probabilities for all five tangles and meshes lie where the simulated RMS excursion is greatest, at intermediate accelerations beyond the slackening threshold, in the
slack--impulsive region (Fig.~\ref{fig:model}B; Supplementary Information section VI). The tension-only (yo-yo-like)  nonlinearity is therefore sufficient for impulse creation. At the highest accelerations, however, the measured  extraction probability declines  even as the simulated RMS excursion remains large, indicating a limit of the lumped point-mass model, which cannot capture the extended size of a physical tangle. This deviation appears  at the estimated transmission crossover (vertical blue band in Fig.~\ref{fig:model}B), which for the stiffness-to-weight ratios $\chi$ of our tangles precedes the model's elastic--inertial boundary. Beyond the crossover, impulses no longer elicit a coherent response from the entire tangle, and the drive's work is absorbed as internal deformation.

Optimal extraction from an amorphous tangle is therefore not a resonance phenomenon. The drive acceleration determines whether the tether can release and recapture the suspended tangle. The  tangle's stiffness (inherited from its topology) sets how far the resulting  tensile response persists. And the transmission time across the extended network then decides whether that response arrives  before the drive reverses. Filament extraction is  maximal where  these local and global requirements overlap. The window's lower edge, $f_{\rm on}=(2\pi)^{-1}\sqrt{g/A}$, depends only on the drive and gravity; its upper edge, from $f_{\pi/2}=v_{\rm eff}/(4L)$ to $f_\pi=v_{\rm eff}/(2L)$, depends only on the tangle. The extraction window can thus be tuned from either side, by the amplitude of the drive or by the size and transmission speed of the tangle.

\section*{Discussion}

Filament extraction requires each drive cycle to complete a sequence of mechanical events. The tangle must first be cohesive enough to carry a localized pull; then the drive must create a slack interval, recapture the freely moving body with sufficient force to initiate slip, and finally communicate that event through the bulk of the tangle before the forcing changes phase. Our tangles and meshes, with measured stiffness-to-weight ratios $\chi\approx200$--$800$, all lie in the same part of the response map where the measured transmission crossover precedes the model's elastic--inertial boundary (Fig.~\ref{fig:model}B). This placement explains why driving faster does not shake the filament free: impulses are still created at the tether, but the tangle can no longer coherently respond to the drive. The two boundaries arise within the same contact network, yet the elastic--inertial boundary limits only how strongly each recapture can load the suspended mass. Whether that load can set the whole tangle in motion is decided by the transmission crossover. This difference is evident in our experiments at high frequency, where the center of mass remains nearly stationary even as individual regions continue to deform (Fig.~\ref{fig:untangling}F). Moreover, because overlapping extraction windows appear in disordered tangles and ordered stitched meshes alike, the competition between creating an impulse and transmitting it depends only weakly on the particular random contact graph. The topology instead contributes the frictional cohesion that each impulse must overcome, and the stiffness and transmission speed that determine how far it propagates.

The drive contributes only  its peak acceleration, which is independent of the  tangle's properties and by itself decides when the tether can fall slack. For a sinusoidally driven suspension, the condition $\Gamma_{\rm acc}=1$ fixes the slackening onset, and increasing the displacement amplitude moves that onset to lower frequency, since $f_{\rm on}\propto A^{-1/2}$. Crossing this boundary permits the tethered filament to unload, but does not guarantee extraction. The subsequent recapture must still overcome the resistance of the embedded frictional path. This distinction is apparent in the weakly cohesive 0-TW tangles, which release a filament over a broad range of frequencies, and in the more cohesive 2- and 10-TW tangles and stitched meshes, for which extraction is narrowly selected. Cohesion (entanglement) thus gives the tangle its ability to bear load, while slack--recapture converts the smooth piston motion into the intermittent events that can overcome that friction. Our model, which contains no friction, therefore predicts where extraction can occur but not its probability. 

Our experiments also leave one region of the response map untested: below $\chi\approx10^{2}$, where the elastic--inertial diagonal crosses the transmission band (Fig.~\ref{fig:model}B), the order of the two boundaries reverses. Extraction from such softer, heavier assemblies should then fail at impulse creation rather than at transmission, with the suspended mass unable to follow the drive at all. Denser filaments or weaker networks would put this prediction to a direct test.

The competition identified here involves only two ingredients, a connector that buckles under compression and a network that carries force at finite speed. It is therefore natural to ask whether the same competition governs any system in which one filament is pulled through a rearranging network of frictional contacts.  In textile assemblies, for example, the resistance to withdrawing one fiber depends on its contacts and interactions with the neighboring fibers~\cite{li1996measurement,qian2024withdrawal}. In a spun yarn, the same frictional contacts transmit the tension that holds cloth together~\cite{warren2018clothes}. 

Agitation, however,  is not intrinsically untangling. Shaking can equally create knots in strings~\cite{raymer2007spontaneous}, destabilize entangled packings~\cite{gravish2012entangled}, or generate connected networks of hooked particles~\cite{kim2025percolation}, so a drive protocol selects one dynamical pathway among many.  Even at the molecular scale, a time-dependent pull can achieve what a steady one cannot, with periodic rather than constant forces triggering knot untying during the translocation of knotted proteins through a pore~\cite{szymczak2016periodic}. In living tangles, finally, the network itself can move. Blackworm blobs  untangle within tens of milliseconds through coordinated body motion~\cite{patil2023ultrafast}, driving their network from within rather than through any single tether. Interestingly,  in our qualitative observation of a freshwater leech (predator) trying to shake one worm free from a blackworm blob (prey), the leech progressively extracts one worm while the the rest of the blob continuously rearranges (Supplementary Video~8), in effect performing our extraction experiment on an active tangle.

When a wire or thread is caught in a tangle,  we often try to shake it loose rather than simply pulling harder. The same practical intuition is passed down in the crafts and the textile industry. Handweaving guides recommend shaking or strumming a warp to help clear snarls~\cite{piroch2004magic,osterkamp2011warp};  chair-seat weaving instructions couple gentle shaking with the withdrawal of individual strands~\cite{comstock1994cane}; and industrial devices vibrate piles of interlocked wire components to isolate them~\cite{kettner1983device}. These practices are not direct realizations of the system studied here, yet they reflect a long-standing empirical intuition that time-dependent forcing can  rearrange a tangled collection of fibers. Our results show that this intuition (shake rather than pull) has an important limitation: more rapid shaking does not necessarily make extraction easier. Each tangle, via its size and transmission speed, encodes its own extraction window. Shaking can thus serve as a dynamical ``sieve'', extracting filaments only from frequency-matched tangles. Since every extraction also reshapes the remaining tangle and shifts its window, controlling the impulse timing could lead to programmable disassembly of tangled matter.

\section*{Acknowledgements}
I.T. thanks Laxminarayanan Krishnan of the Georgia Tech Microcomputed Tomography Core for training and assistance with acquiring the microcomputed tomography scans, Materials Innovation and Learning Laboratory (The MILL) at Georgia Tech for training and usage of the Mark-10 mechanical tester, Harry Tuazon for sharing the worm--leech interaction video, Carlos Baez for helpful discussions about instance segmentation of $\mu$CT data, and Nitesh Arora for discussions and for prototyping a larger-scale untangling apparatus for follow-on experiments with handling heavier filaments. I.T. and Saad Bhamla also thank the members of the Bhamla lab at Georgia Tech for constructive discussions. 
\subsection*{Funding}
S.B. acknowledges funding support from NSF Grant PHY-2310691 and Schmidt Sciences. This research received support through Schmidt Sciences, LLC (to J.D.), and the MathWorks Professorship Fund (to J.D.).

\subsection*{Artificial Intelligence disclosure}
Large language models (ChatGPT and Claude) were used to refine the language of the manuscript and to generate comments in the analysis code. The authors reviewed and take full responsibility for all content.\\
\section*{Author contributions}
Ishant Tiwari, Vishal Patil and Saad Bhamla conceived and designed the study. Ishant Tiwari designed and performed experiments and developed the numerical model and analysis pipelines. Sutikshan Bansal performed the frequency-scanning experiments. Ishant Tiwari and Sutikshan Bansal manually instance-segmented the micro-CT data. Rosalind Huang, Chenyi Fei, and J\"orn Dunkel developed the machine-learning-based parameter-inference methodology. Saad Bhamla and J\"orn Dunkel supervised the project and contributed to experimental design and interpretation. Vishal Patil developed the topology analysis pipelines. Ishant Tiwari and Saad Bhamla wrote the initial draft. All authors reviewed and edited the manuscript.\\
\section*{Competing Interests}
The authors declare no competing interests.\\
\section*{Code Availability}
The analysed data from the frequency scan videos, the micro-CT scans, 2D projected tangles and the simulations along with the analysis codes used for this paper are available on the github link~\cite{github}. Raw video files for individual experiments are available from the authors upon request.

\bibliography{references}

\end{document}


\title{Supplementary Information: Why shaking faster does not untangle: frequency-selective extraction of filaments from tangled matter}
\author{Ishant Tiwari}
\email[Correspondence: ]{ishant@coloradocollege.edu}
\affiliation{Department of Physics, Colorado College, Colorado Springs, Colorado 80903, USA}
\affiliation{School of Chemical and Biomolecular Engineering, Georgia Institute of Technology, Atlanta, Georgia 30332, USA}

\author{Sutikshan Bansal}
\thanks{These authors contributed equally.}
\affiliation{School of Materials Science and Engineering, Georgia Institute of Technology, Atlanta, Georgia 30332, USA}

\author{Rosalind Huang}
\thanks{These authors contributed equally.}
\affiliation{Department of Physics, Massachusetts Institute of Technology, Cambridge, Massachusetts 02139, USA}

\author{Chenyi Fei}
\affiliation{Department of Mathematics, Massachusetts Institute of Technology, Cambridge, Massachusetts 02139, USA}

\author{J\"orn Dunkel}
\affiliation{Department of Mathematics, Massachusetts Institute of Technology, Cambridge, Massachusetts 02139, USA}

\author{Vishal Patil}
\affiliation{Department of Mathematics, University of California San Diego, La Jolla, California 92093, USA}

\author{Saad Bhamla}
\email[Correspondence: ]{saad.bhamla@colorado.edu}
\affiliation{BioFrontiers Institute and Department of Chemical and Biological Engineering, University of Colorado Boulder, Boulder, Colorado 80309, USA}
\maketitle

\section{Preparation of samples}

\subsection{Stochastic tangles}

As mentioned in the main manuscript, each tangle was prepared from an ensemble of 20 cotton threads. The threads were cut from a cotton spool of diameter approximately $200~\mu$m. Each thread had length $10$ cm. Since the threads retain a finite curvature after being unwound from the spool, we first crumpled each thread manually and then stretched it again before using it in the experiment. This procedure helped remove the memory of the spool curvature and made the initial state of the different threads more homogeneous.

The prepared threads were placed inside a cuboidal tumbling box of dimensions approximately $10~{\rm cm}\times 10~{\rm cm}\times 5~{\rm cm}$. The depth of the box was $5$ cm. Steel nuts were used as tumbling weights (TWs) to agitate the threads during rotation. We used M3 nuts as TWs, and varied their number as $N_{\rm TW}=0,2,10$ or $20$. After placing the threads and TWs inside the box, the box was closed and attached to a stepper motor such that the rotation axis passed through the centre of the box along its depth direction.

The stepper motor was controlled using an Arduino. The box was rotated clockwise for $2$ s and then counterclockwise for $2$ s, with the angular speed of each rotation chosen randomly by the Arduino. This clockwise--counterclockwise driving cycle was repeated for a total duration of $2$ min. The repeated tumbling motion caused the initially separated threads to bend, collide and intertwine with each other, resulting in a compact tangled ball. This preparation protocol was used for all samples, unless stated otherwise, and produced tangles that were reproducible enough for the subsequent measurements. The complete stochastic-tangle preparation sequence is shown in Supplementary Video~1.

\subsection{Deterministic meshes}

To construct deterministic tangled samples, we used cotton gauze sponges (Curity). The original gauze sponge had dimensions $2~{\rm inch}\times 2~{\rm inch}$ and consisted of multiple sheets of woven cotton mesh. A single sheet was separated from the sponge and cut to obtain a sample mass comparable to that of the stochastic tangled samples, typically $40$--$60$ mg. The resulting sample was a square-grid mesh of overlapping cotton threads.

The deterministic sample was suspended using a single tethering thread. This thread was passed through the holes of the gauze mesh and then attached to the vertically oscillating piston. The piston drove the sample by moving up and down, while the gauze sheet provided a prescribed and reproducible contact geometry between the tether and the mesh.

\begin{figure}[h]
    \centering
    \includegraphics[width=0.5\linewidth]{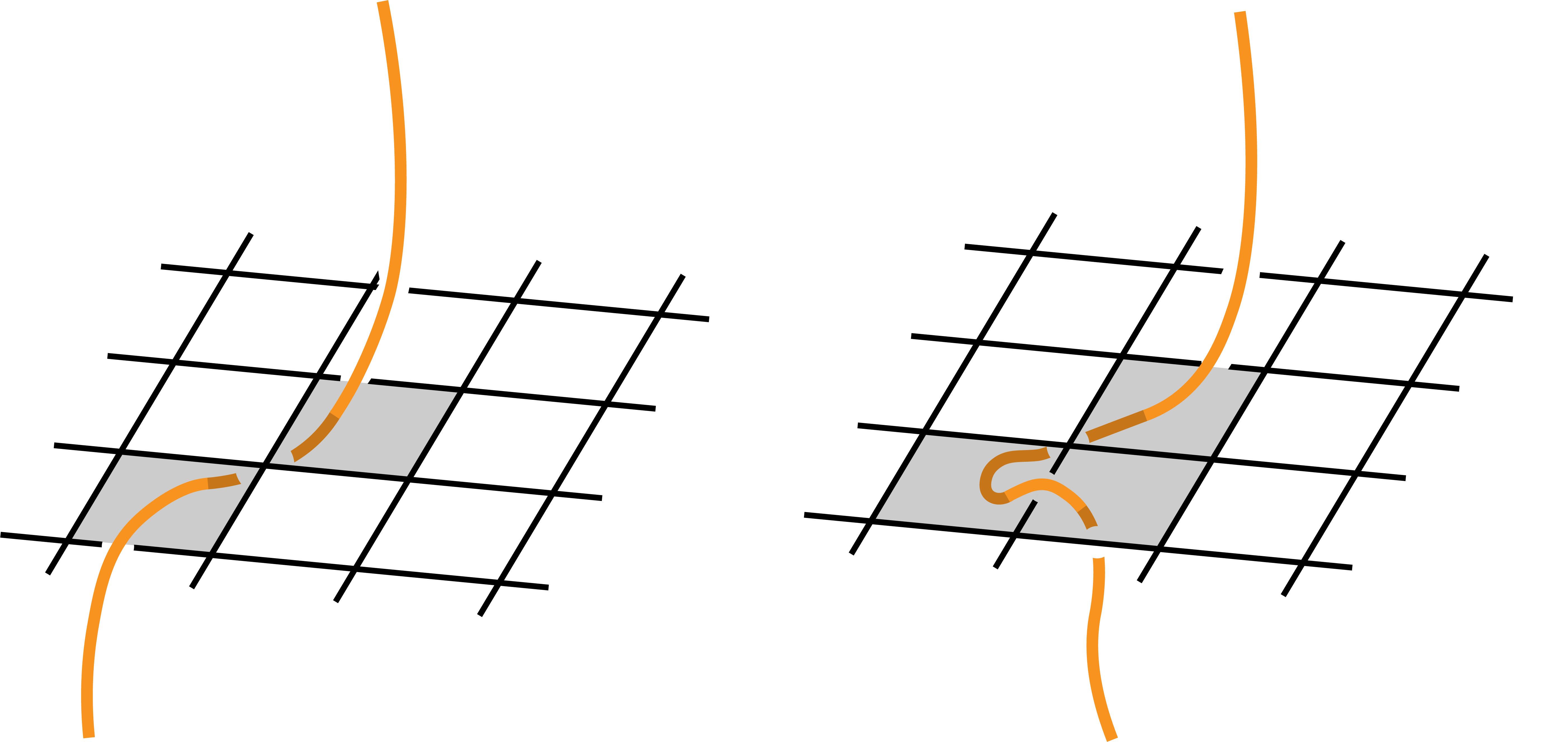}
    \caption{Stitching pattern for $s=2$ and $s=3$ stitches through a single sheet of cotton gauze.}
    \label{fig:deterministic_stitch}
\end{figure}

To tune the effective degree of entanglement between the tethering thread and the mesh, we varied the number of stitches, denoted by $s$. For $s=2$, the tethering thread was passed through the mesh twice, forming one contact loop with the gauze. For $s=3$, the tethering thread was passed through the mesh three times, producing an additional contact with the gauze. The stitching protocols are shown schematically in SI Fig.~\ref{fig:deterministic_stitch}. %

Thus, increasing $s$ increases the number of frictional contacts between the tethering thread and the deterministic mesh. This construction allowed us to compare samples with the same underlying ordered mesh geometry, but with different levels of tether--mesh coupling.

\section{Filament-extraction apparatus}

\begin{figure}[h]
    \centering
    \includegraphics[width=0.7\linewidth]{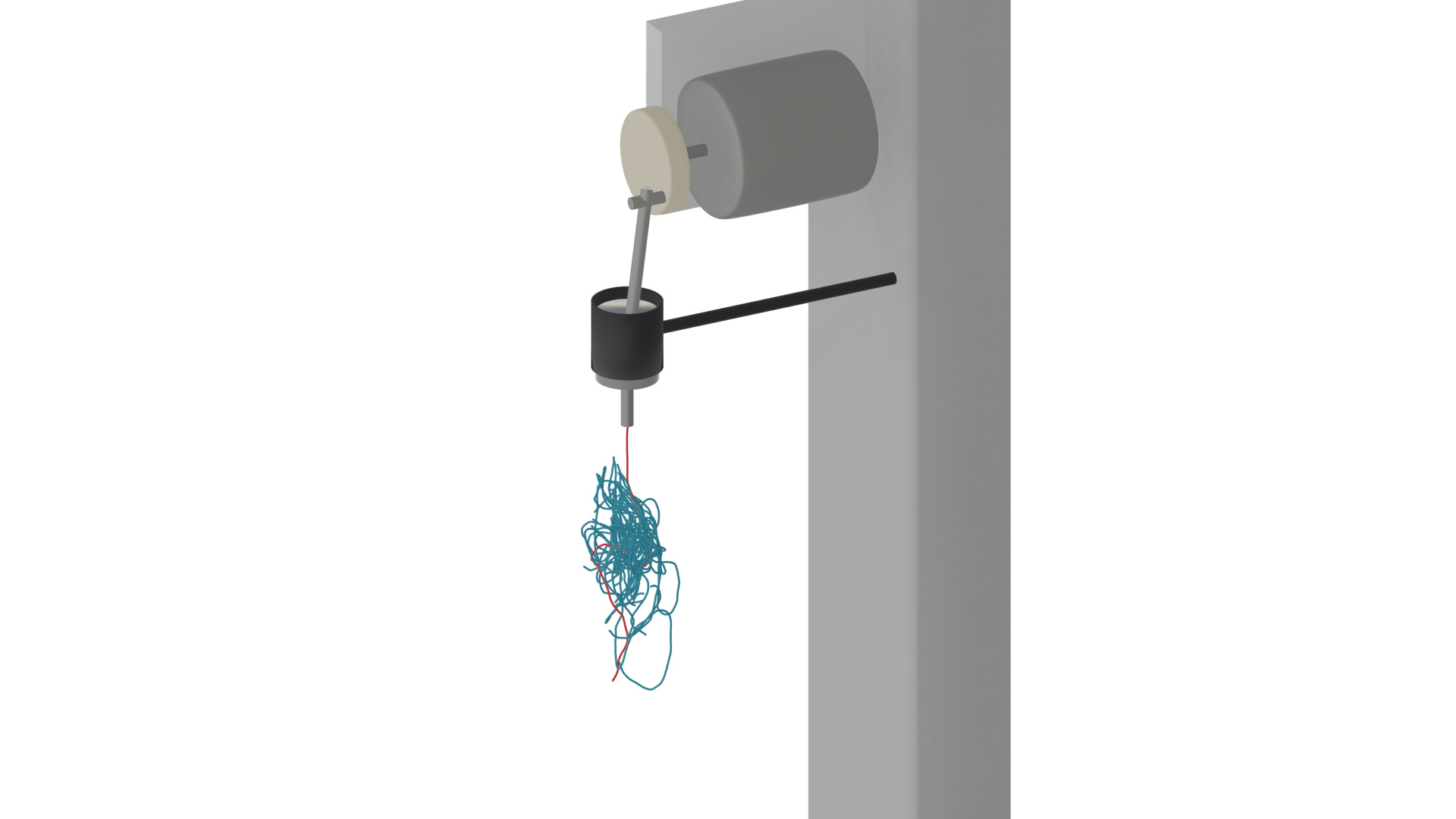}
    \caption{Schematic of the filament-extraction apparatus.}
    \label{fig:untangler}
\end{figure}

To extract a thread from either the stochastic or deterministic samples, we used a vertically oscillating piston driven by a rotary DC motor (SI Fig.~\ref{fig:untangler}). A wheel--piston--sleeve assembly converted rotation into approximately sinusoidal vertical motion. The measured peak-to-peak stroke was approximately $1.5\,$cm, corresponding to a one-sided amplitude $A=0.75\,$cm. Thus, varying frequency at fixed stroke also varied the maximum support speed $2\pi fA$ and the dimensionless maximum support acceleration
\begin{equation}
    \Gamma_{\rm acc}=\frac{A(2\pi f)^2}{g}.
\end{equation}
For the reported amplitude, $\Gamma_{\rm acc}=1$ at $f=5.8\,$Hz. The motor frequency was measured for every trial rather than assigned from the nominal voltage. The pooled distribution across sample types is summarized below; all response plots use the measured trial frequencies (or their condition-wise medians), not nominal set points.

\begin{table}[h]
\centering
\caption{Measured drive frequencies pooled across the seven stochastic-tangle and mesh conditions at each of the five nominal drive levels.}
\begin{tabular}{ccccc}
\hline
Level & mean (Hz) & SD (Hz) & range (Hz) & median (Hz) \\
\hline
1 & 5.03 & 1.12 & 3.23--7.13 & 5.08 \\
2 & 8.96 & 1.59 & 5.24--12.88 & 8.94 \\
3 & 14.86 & 1.74 & 8.99--18.85 & 15.08 \\
4 & 24.13 & 1.78 & 20.31--27.74 & 24.50 \\
5 & 36.30 & 1.52 & 31.22--39.80 & 36.40 \\
\hline
\end{tabular}%
\label{tab:measured_frequency}
\end{table}

The oscillation frequency was controlled by varying the voltage supplied to the DC motor. Increasing the voltage increased the rotational frequency of the motor and, consequently, the vertical oscillation frequency of the piston. In our experiments, we varied the oscillation frequency over a range of approximately $4$--$40$ Hz, while maintaining the same peak-to-peak displacement of the nail.

For stochastic tangled samples, one of the 20 threads forming the tangle was tethered to the nail at the bottom of the piston. The remaining threads in the tangled ball were left freely suspended below the attachment point. During shaking, the piston therefore pulled periodically on a single thread, while the rest of the tangle moved in response to the imposed oscillatory forcing. For deterministic samples, the tethering thread that was stitched through the gauze mesh was tied directly to the nail. The same piston motion was then used to drive the deterministic mesh through the prescribed stitch geometry.

\section{X-ray computed tomography and instance segmentation}

The three-dimensional structure of the cotton tangles was measured using X-ray computed tomography (XCT). The scans were performed with a Scanco $\mu$CT50 micro-CT scanner. The reconstructed image slices had a resolution of $2048 \times 2048$ pixels, with a voxel size of $17.2~\mu$m. The scanner was operated at a peak voltage of $45$ kV, a current of $88~\mu$A and an exposure time of $500$ ms per projection.

For imaging, the cotton tangles were suspended inside a cylindrical sample holder of diameter $34$ mm. To hold the tangle in place without compressing it, two to three free thread ends were taped to the roof of the sample holder, allowing the rest of the tangle to hang freely. During imaging, the sample holder was rotated while the scanner acquired the transmitted X-ray data. The tomography data were reconstructed into a stack of two-dimensional grayscale image slices along the $z$ direction using the proprietary Scanco evaluation software. The reconstructed image stacks were then exported in DICOM format for subsequent analysis.

The first step in the image-processing pipeline was semantic segmentation, in which the grayscale XCT images were separated into foreground voxels belonging to cotton threads and background voxels corresponding to empty space. This semantic segmentation was performed in Dragonfly workstation software (Comet Group) using a U-Net convolutional neural network. The U-Net takes a single-channel two-dimensional grayscale image as input and returns a pixel-wise probability map. For the binary segmentation used here, each pixel was assigned two probabilities, corresponding to the cotton-thread foreground class and the background class. The final semantic segmentation was obtained by assigning each pixel to the class with the larger predicted probability.

After semantic segmentation, we manually performed instance segmentation of the cotton threads. In contrast to semantic segmentation, which only distinguishes cotton from background, instance segmentation assigns each foreground voxel to a specific thread identity. This step was performed by visually inspecting the three-dimensional XCT image stack in Dragonfly and manually painting the voxels associated with each thread. Voxels belonging to the first identified thread were assigned one label, voxels belonging to a second thread were assigned another label, and so on until the individual threads in the tangle were separated.

This manual instance-segmentation procedure was laborious and time-consuming, but it avoided ambiguities that can arise in automated segmentation of densely contacting or overlapping threads. In particular, manual inspection allowed us to use local continuity, grayscale contrast and the three-dimensional geometry of each fiber to decide thread identity near contact points. The resulting labelled image stacks therefore provided a reliable voxel-level representation of the individual threads, which was used for subsequent geometrical and topological analysis of the tangles.

The projected size measures in main-text Fig.~1E--F use calibrated tangle images (118.54~pixels/mm). We analyzed three independently prepared tangles per condition (0, 2 and 10 TW). A morphological-gradient mask was closed and hole-filled, the largest connected tangle region was retained, and foreground pixels within it were selected by an intensity threshold of 0.20. For foreground positions $\mathbf p_i$ and centroid $\bar{\mathbf p}$, the population covariance
\begin{equation}
    \mathbf C=\frac{1}{N}\sum_{i=1}^{N}
    (\mathbf p_i-\bar{\mathbf p})(\mathbf p_i-\bar{\mathbf p})^{\mathsf T}
\end{equation}
has eigenvalues $\lambda_1\geq\lambda_2$. The plotted major- and minor-axis RMS extents are $\sqrt{\lambda_1}/s_{\rm img}$ and $\sqrt{\lambda_2}/s_{\rm img}$, where $s_{\rm img}$ is the pixel calibration. Principal-axis measures are insensitive to in-plane image rotation but depend on out-of-plane viewing angle. The mean major-axis extents for 0, 2 and 10 TW are 9.01, 7.43 and 6.55~mm; the corresponding minor-axis extents are 6.21, 5.43 and 5.03~mm.
We also calculated $R_{95}$, the 95th percentile of foreground-pixel distance from the centroid, and used $L=2R_{95}$ as a projected edge-to-edge traversal-length proxy. For the three 2-TW and three 10-TW images, the pooled result is $L_{95}=27\pm1.8\,$mm (mean $\pm$ sample standard deviation). The condition means are 28.2~mm (2 TW) and 25.9~mm (10 TW).
For each representative micro-CT reconstruction, we calculated the occupied envelope of the traced centerlines using a three-dimensional alpha shape with a common radius $\alpha=7.00\,$mm. The envelope volumes are 12.2, 7.6 and 7.0~cm$^3$ for 0, 2 and 10 TW. The corresponding equivalent-sphere diameter, $D_{\rm eq}=(6V/\pi)^{1/3}$, is 28.6, 24.4 and 23.7~mm. Each condition contains one 3D reconstructed tangle.

For the thread-level contact graphs in main-text Fig.~1G, each of the 20 manually traced centerlines was treated as one node. Two threads were assigned an undirected edge when the minimum distance between their piecewise-linear centerlines was less than $0.25\,$mm, the approximate thread diameter used in the analysis. The minimum distance was evaluated between line segments, rather than only between the sampled vertices. Multiple contact regions between the same pair were collapsed to one edge. This gave 26, 49, and 79 contacting pairs in the representative 0-, 2-, and 10-TW reconstructions, respectively.

We quantified the crossing complexity of the open centerlines using the pairwise average crossing number (ACN). For threads $i$ and $j$,
\begin{equation}
    {\rm ACN}_{ij}
    =\frac{1}{2\pi}
    \int_{C_i}\int_{C_j}
    \frac{\left| (d\mathbf r_i\times d\mathbf r_j)\cdot
    (\mathbf r_i-\mathbf r_j)\right|}
    {|\mathbf r_i-\mathbf r_j|^3}.
    \label{eq:si_pair_acn}
\end{equation}
This is the average unsigned crossing count over projection directions, following the pairwise definition of Becker \emph{et al.}~\cite{becker2022active}. The value plotted for thread $i$ in main-text Fig.~1H is its ACN strength, $a_i=\sum_{j\ne i}{\rm ACN}_{ij}$. The mean values across the 20 threads were 90.10, 153.52, and 189.81 for 0, 2, and 10 TW, respectively.

We also calculated a contact-link measure that retains mutual winding only for pairs that are close enough to be in physical contact. The signed open-curve Gauss integral is
\begin{equation}
    L_{ij}
    =\frac{1}{4\pi}
    \int_{C_i}\int_{C_j}
    \frac{(d\mathbf r_i\times d\mathbf r_j)\cdot
    (\mathbf r_i-\mathbf r_j)}
    {|\mathbf r_i-\mathbf r_j|^3}.
    \label{eq:si_open_gauss_link}
\end{equation}
If $A_{ij}$ is the binary contact matrix defined above, the contact-link value for thread $i$ is
\begin{equation}
    c_i=\sum_{j\ne i}A_{ij}|L_{ij}|.
    \label{eq:si_contact_link}
\end{equation}
The distribution of $c_i$ is shown in main-text Fig.~1I. Its mean across the 20 threads increased from 1.1 to 2.4 and 5.1 for 0, 2, and 10 TW. In the network drawings, every detected physical contact is retained, while line shade and width scale with the square root of $|L_{ij}|$ normalized by the largest value across the three reconstructions.

The Gauss integrals were evaluated by the exact solid-angle formula for pairs of polygonal segments after resampling each centerline uniformly in arc length. We used 250 points per thread for the plotted values and repeated the calculation with 400 points per thread as a numerical check. The condition-mean ACN changed by less than $0.75\%$, and the condition-mean contact link by less than $0.007\%$, between these two resolutions. Because the reconstructed threads are open, $L_{ij}$ is generally not an integer and is not a strict topological linking invariant. Here it is used as a geometric winding descriptor conditioned on physical contact. Finally, there is one reconstructed tangle per TW condition. The 20 thread-level values in each box plot are subsamples within that reconstruction, not independent tangle replicates.

\section{Quasi-static extraction experiments}

To measure the resistance of the tangles to direct thread extraction, we performed quasi-static pulling experiments using a Mark-10 FS05 mechanical testing device equipped with a $0.5$ N force sensor. Stochastic tangles were prepared using the same tumbling protocol described above, with $N_{\rm TW}=0,2$ or $10$ tumbling weights. For each value of $N_{\rm TW}$, we tested three independently prepared tangles. In each measurement, one thread was selected from the tangle and tethered to the force sensor, while the rest of the tangle was left suspended below it.

\begin{figure}[h]
    \centering
    \includegraphics[width=0.7\linewidth]{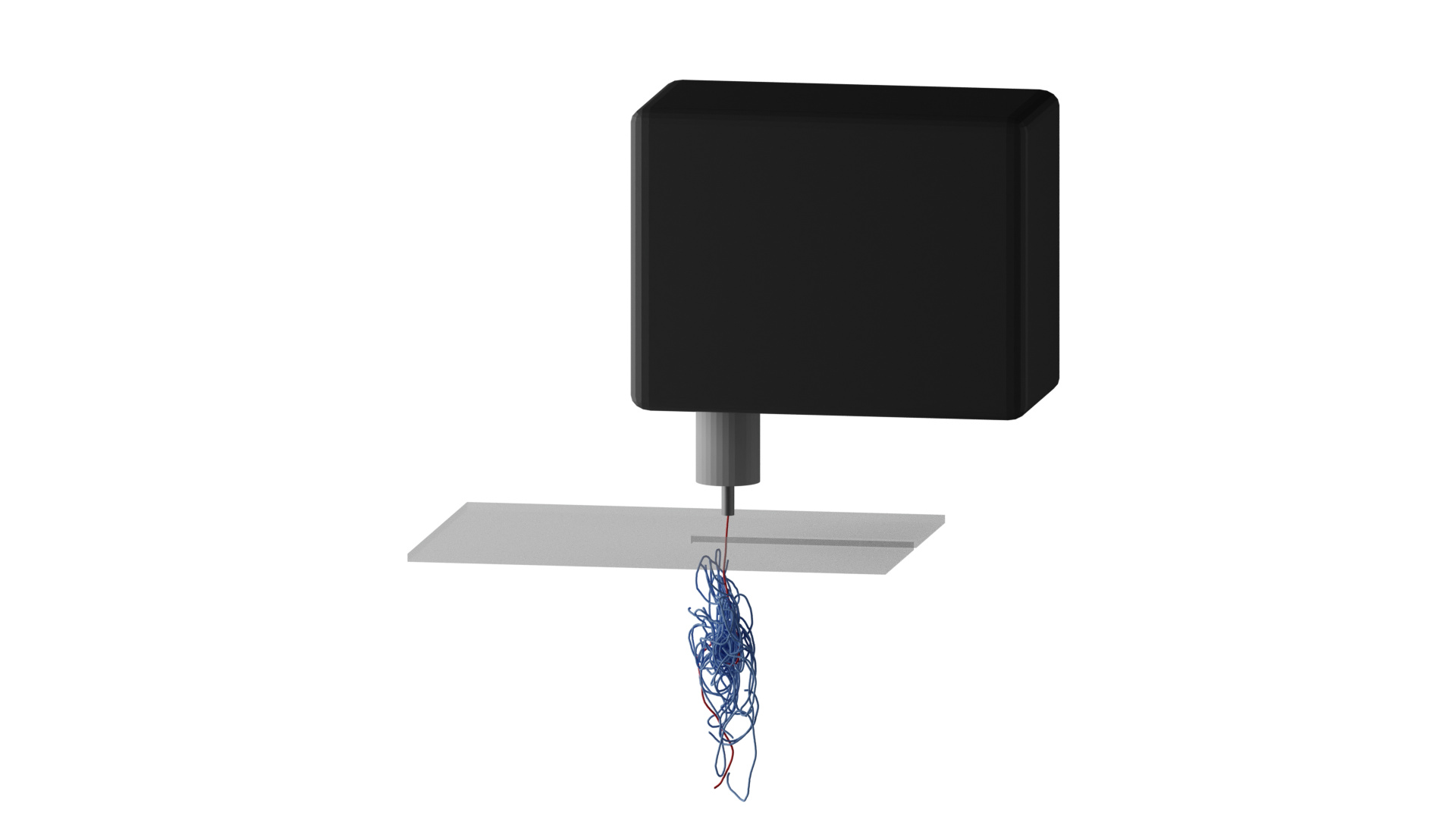}
    \caption{Schematic of the quasistatic extraction setup.}
    \label{fig:quasi_schematic}
\end{figure}

To ensure that the measured force corresponded to pulling one thread through the surrounding tangle, we placed a rigid slit between the force sensor and the tangle (Fig.~\ref{fig:quasi_schematic}). The tethered thread passed through this slit, while the remaining tangled aggregate was held below it. The slit width was larger than the thread diameter, but much smaller than the tangle size, so that the pulled thread could pass through while the tangle body was blocked. As the mechanical tester moved the force sensor upward, the tethered thread was pulled through the tangle. The pulling speed was fixed at $15~{\rm mm\,min^{-1}}$ for all measurements.

The resulting force--displacement curves showed a characteristic stick--slip response. During sticking intervals, the measured tensile force increased as the pulled thread loaded the surrounding tangle. This increase was followed by sharp force drops, corresponding to slip events in which the pulled thread moved through one or more frictional contacts. The maximum tensile force measured during each extraction curve was used in the main text to quantify the quasi-static extraction force.

We also used these force--displacement curves to estimate an effective stiffness associated with the loading portions of the stick--slip response. Before analysis, the force signal was baseline-corrected using the initial unloaded portion of the measurement, and the displacement was redefined relative to the point at which the thread first became mechanically engaged with the tangle. The force--displacement data were then resampled onto a uniform displacement grid.

For each extraction curve, we identified stick--slip loading intervals as regions in which the force increased from a local minimum to the subsequent local maximum. These local maxima were detected using a peak-prominence criterion, and the preceding local minimum was used to define the start of the corresponding loading interval. For each accepted interval, we fit the force--displacement data to a linear relation
\begin{equation}
    F = F_0 + k_{\rm eff} x,
\end{equation}
where $F$ is the measured tensile force, $x$ is the engaged displacement, $F_0$ is a fitted intercept and $k_{\rm eff}$ is the fitted slope. Thus, the stiffness was obtained from a linear fit to the data points within each loading interval. Points associated with small internal force drops within a loading interval were excluded from the fit when detected.

This procedure yielded a set of effective stiffness values for the loading intervals in each extraction curve. We used the median of these interval-wise fitted stiffnesses as the representative stiffness for that tangle. This median-based estimate reduces sensitivity to occasional poorly resolved slip events or unusually short loading intervals.

\begin{figure}[h]
    \centering
    \includegraphics[width=\linewidth]{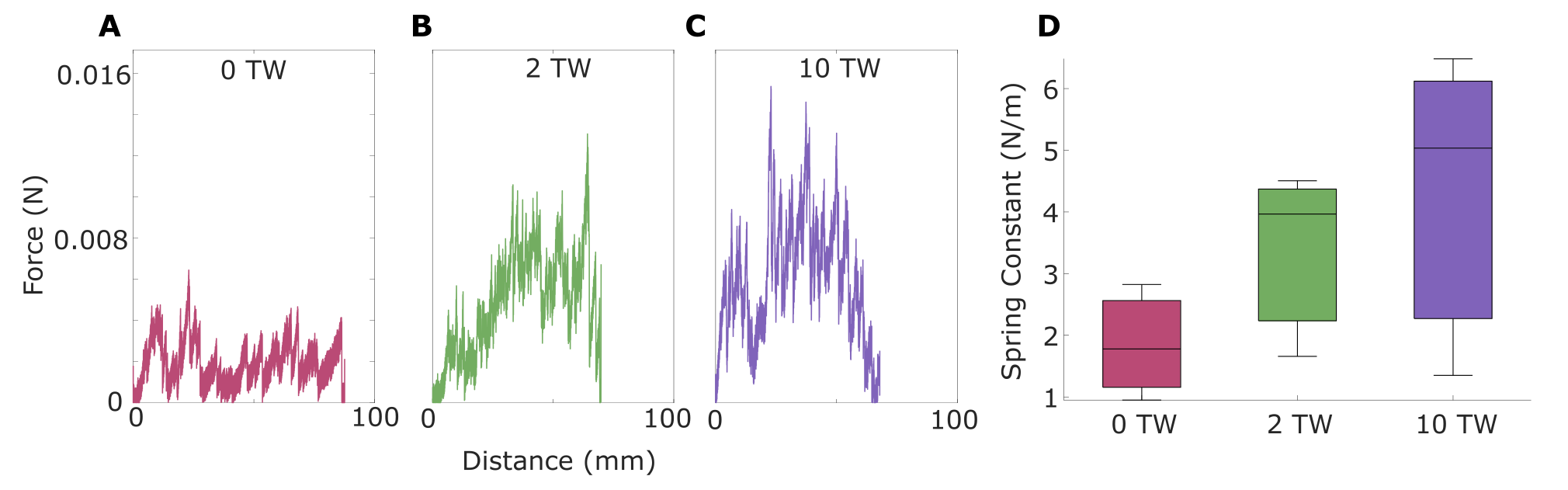}
    \caption{A-C) Force as a function of displacement of the tethered thread in the quasistatic pullout experiments for the stochastic tangle made with A) 0 TW B) 2 TW C) 10 TW. D) Measured spring constant/stiffness of the tangle from epochs of monotonic force increase in the force vs. displacement curves. The box boundaries indicate upper and lower quartiles, while the line inside the box represents the median. The whiskers on the boxes represent the lowest and highest values of the distribution.}
    \label{fig:quasi_tangle_data}
\end{figure}

This stiffness is an effective, quasi-static measure of how rapidly the tangle builds tension before a slip event. It is not a microscopic spring constant of an individual thread. Rather, it reflects the collective compliance of the pulled thread, the surrounding fiber contacts and the local rearrangements that precede slip. The resulting stiffness values for tangles prepared with $N_{\rm TW}=0,2$ and $10$ are shown in Fig.~\ref{fig:quasi_tangle_data}.

We performed the same quasi-static extraction analysis on the deterministic gauze meshes. In this case, the sample was prepared with a prescribed stitch number, $s=2$ or $s=3$, and the stitched tethering thread was attached to the force sensor. The mesh body was held below the slit in the same manner as the stochastic tangles, so that upward motion of the sensor pulled the tethering thread through the prescribed mesh contacts. This geometry allowed us to measure the resistance associated with a controlled number of thread--mesh contacts, without the additional structural disorder present in the stochastic tangles.

\begin{figure}[h]
    \centering
    \includegraphics[width=0.9\linewidth]{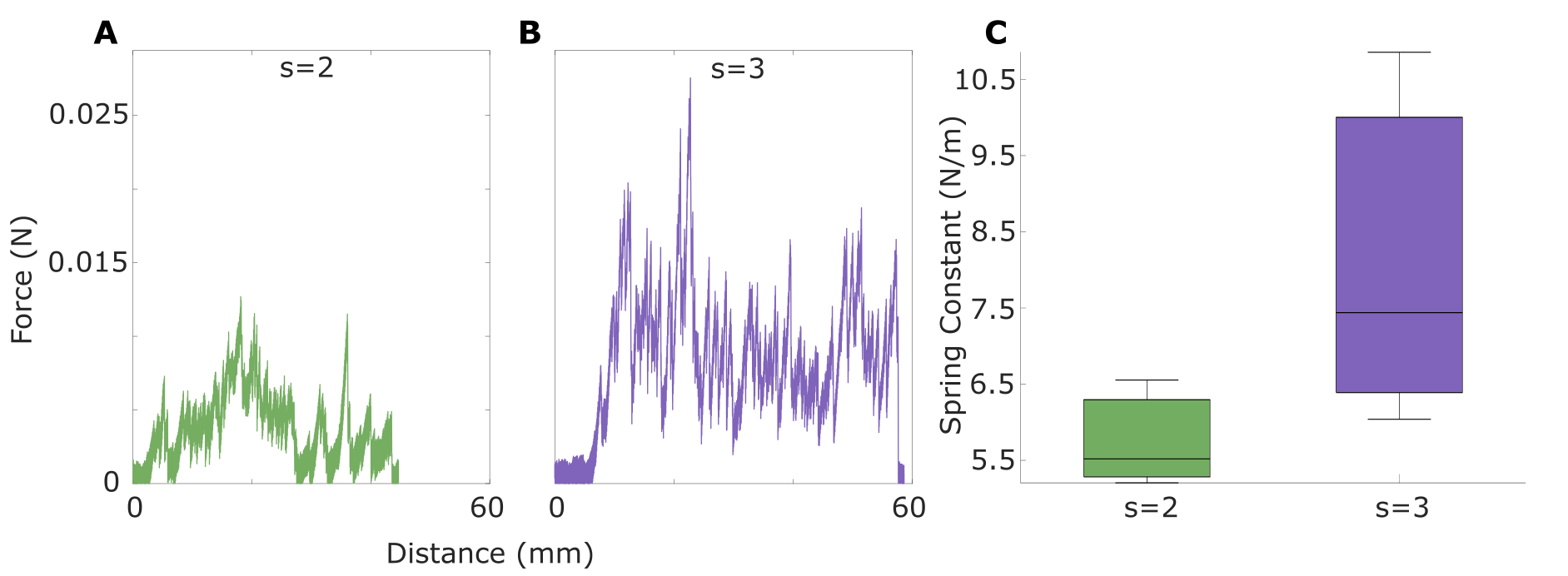}
    \caption{A,B) Force as a function of displacement of the tethered thread in the quasistatic pullout experiments for the mesh with A) $s=2$ and B) $s=3$ stitches. C) Measured spring constant/stiffness of the mesh from epochs of monotonic force increase in the force vs. displacement curves. The box boundaries indicate upper and lower quartiles, while the line inside the box represents the median. The whiskers on the boxes represent the lowest and highest values of the distribution.}
    \label{fig:quasi_mesh_data}
\end{figure}

The deterministic meshes also exhibited stick--slip force--displacement curves. We analyzed these curves using the same procedure as for the stochastic tangles: loading intervals were identified between local minima and subsequent local maxima, and an effective stiffness was obtained from the slope of the force increase during each interval. The resulting stiffness distributions for $s=2$ and $s=3$ are shown in Fig.~\ref{fig:quasi_mesh_data}. This comparison provides the quasi-static counterpart to the dynamic deterministic-mesh experiments, and shows how increasing the prescribed stitch number changes the force buildup before slip.

The quasi-static $k_{\rm eff}$ values for the 0-, 2-, and 10-TW tangles and the $s=2$ and $s=3$ meshes are used to position the corresponding experimental markers on the main-text response map (main-text Fig.~4B).

\section{Filament-extraction experiments}

\subsection{Recording protocol}

For each filament-extraction experiment, the sample was suspended from the nail attached to the bottom of the vertically oscillating piston. For stochastic tangles, one thread out of the 20 threads in the tangle was chosen at random and tied to the nail. The rest of the tangle was left freely suspended below the attachment point. For deterministic gauze samples, the tethering thread used to stitch through the mesh was tied directly to the nail. After the sample was attached, the DC motor was switched on and the piston was driven vertically at the prescribed frequency.

Each trial was run for a total duration of $300$~s. A successful event was defined as complete extraction of the tethered thread, causing the remaining tangle or mesh to detach. Samples still attached at 300~s were treated as right-censored failures for the binary endpoint. At every preparation--frequency point, $n=10$ independent trials were performed. We report $P_{\rm ext}=k/n$, where $k$ is the number extracted within 300~s, together with two-sided 95\% Wilson binomial confidence intervals. This endpoint is a finite-window extraction probability, not an extraction rate. For successful trials, extraction time was measured from motor onset to detachment.

Two types of video recordings were used. To reconstruct the three-dimensional motion of the tangle center of mass, selected experiments were recorded simultaneously from two orthogonal viewing directions using two Chronos high-speed cameras. Both cameras recorded at $250$ frames per second. These recordings were used to track the center-of-mass trajectory and quantify the spatial excursions and instantaneous speeds of the driven tangle.

For the extraction-probability measurements, each experiment was recorded from a single viewing direction. A Chronos high-speed camera recorded the first $60$ s of the trial at $250$ frames per second, allowing us to resolve the early-time dynamics and rapid extraction events. In parallel, a GoPro camera recorded the full $300$ s duration of the trial at $240$ frames per second. The GoPro recording provided a redundant full-duration record, which was used to determine whether extraction occurred within the complete experimental window and, when applicable, to measure the corresponding extraction time. Supplementary Video~2 shows a representative filament-extraction trial and the repeated tensioning and slip events that precede detachment.

\subsection{Stochastic tangles}

\begin{figure}[h]
    \centering
    \includegraphics[width=0.8\linewidth]{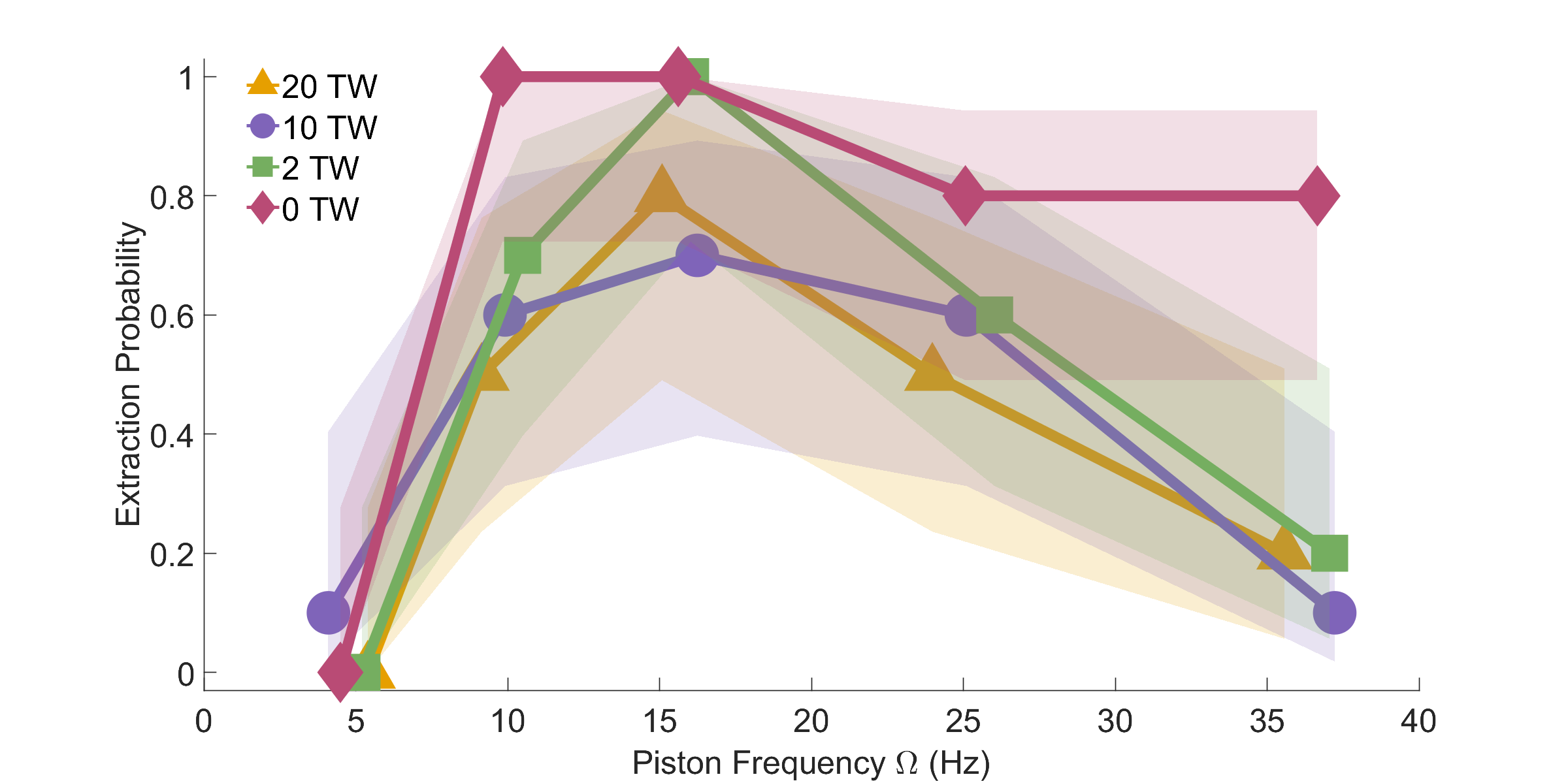}
    \caption{Probability of extraction within 300~s for stochastic tangles prepared with 0, 2, 10, or 20 TW. Each point contains $n=10$ trials; shaded bands are 95\% Wilson binomial intervals.}
    \label{fig:stochastic_tangle}
\end{figure}

For completeness, we also measured the frequency-dependent extraction probability for stochastic tangles prepared with $N_{\rm TW}=20$ tumbling weights. The $20$ TW samples showed the same qualitative behavior as the $10$ TW samples (Fig.~\ref{fig:stochastic_tangle}). At low driving frequency, the aggregate largely followed the piston motion and thread extraction was inefficient. At intermediate frequency, the tangle underwent large, intermittent excursions, producing repeated tensioning and slipping events at the pulled thread. This regime gave the largest extraction probability. At high frequency, the center-of-mass motion became localized and the extraction probability decreased again.

The 20-TW data show the same broad intermediate-frequency maximum as the 10-TW condition, within wide binomial uncertainty. These measurements establish nonmonotonicity over the sampled range but do not locate a unique sharp optimum.

\subsection{Deterministic meshes}

\begin{figure}[h]
    \centering
    \includegraphics[width=0.8\linewidth]{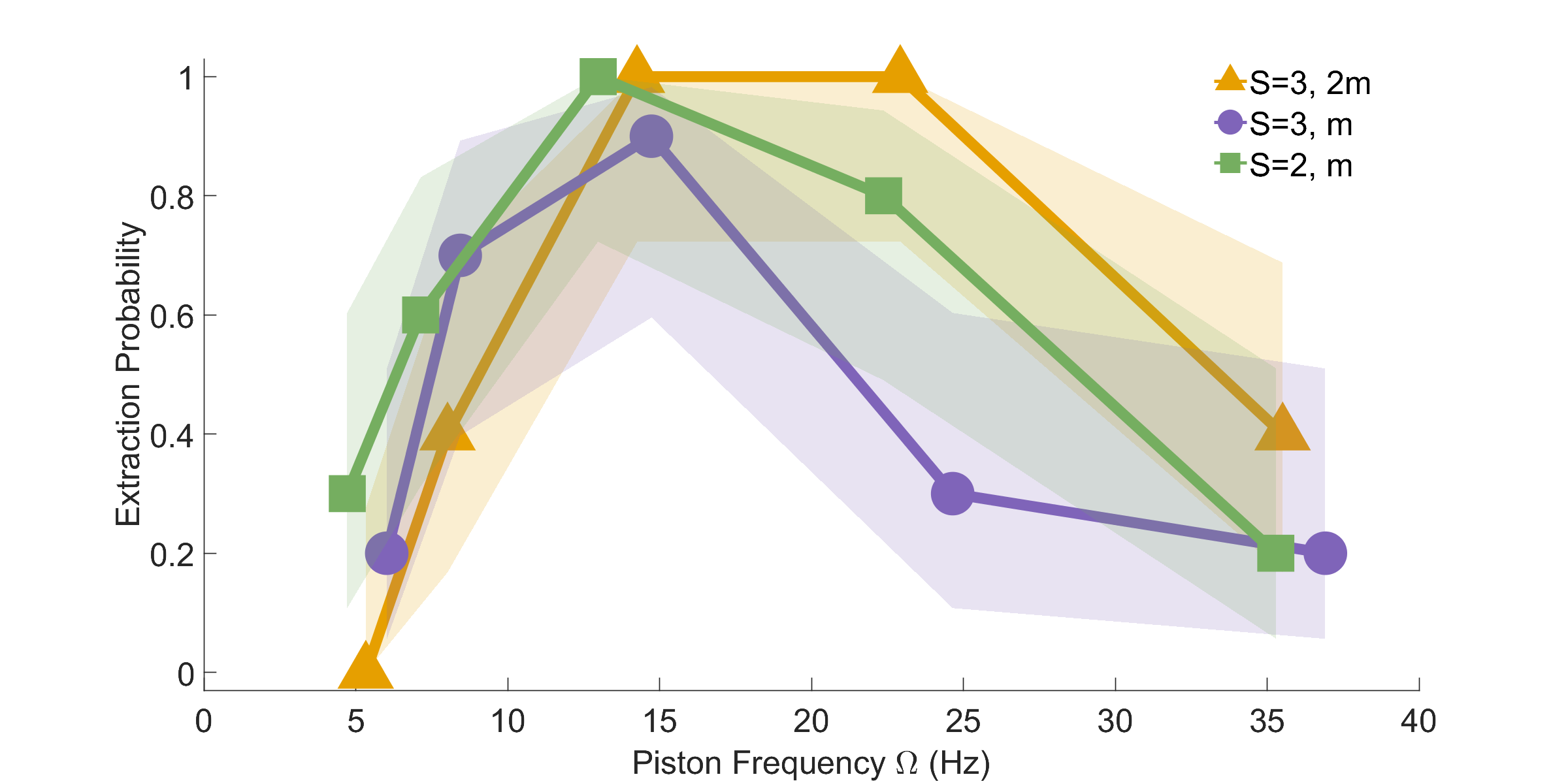}
    \caption{Probability of extraction within 300~s for deterministic meshes with $s=2$ and $s=3$ at mass $m$, and $s=3$ at approximately $2m$. Each point contains $n=10$ trials; shaded bands are 95\% Wilson binomial intervals.}
    \label{fig:deterministic_mesh}
\end{figure}

In the main text, we report the frequency-dependent extraction probability for deterministic gauze samples prepared with stitch numbers $s=2$ and $s=3$. For completeness, we also performed the same filament-extraction experiments on an additional deterministic sample with $s=3$, but with approximately twice the mass of the $s=3$ sample shown in the main text (Fig.~\ref{fig:deterministic_mesh}). The stitch geometry was kept fixed, so that the number of tether--mesh contacts was unchanged, while the total suspended mass of the gauze was increased.

The heavier $s=3$ sample showed the same qualitative frequency-dependent response as the lighter deterministic mesh. At low driving frequency, extraction remained inefficient because the mesh largely followed the imposed piston motion. At intermediate frequency, the mesh underwent larger intermittent excursions, producing repeated tensioning and slipping events at the stitched tether. This regime again gave the highest extraction probability. At high frequency, the motion became more localized, although the decrease in extraction probability was less pronounced than for the lighter $s=3$ sample.

The doubled-mass $s=3$ control retains high extraction probability over a broader part of the sampled intermediate-to-high frequency range than the lighter $s=3$ sample. This difference is consistent with a role for inertial loading, but the two mass conditions alone do not establish a scaling law or isolate mass from sample-to-sample differences.

\subsection{Time to extraction}

\begin{figure}[h]
    \centering
    \includegraphics[width=0.92\linewidth]{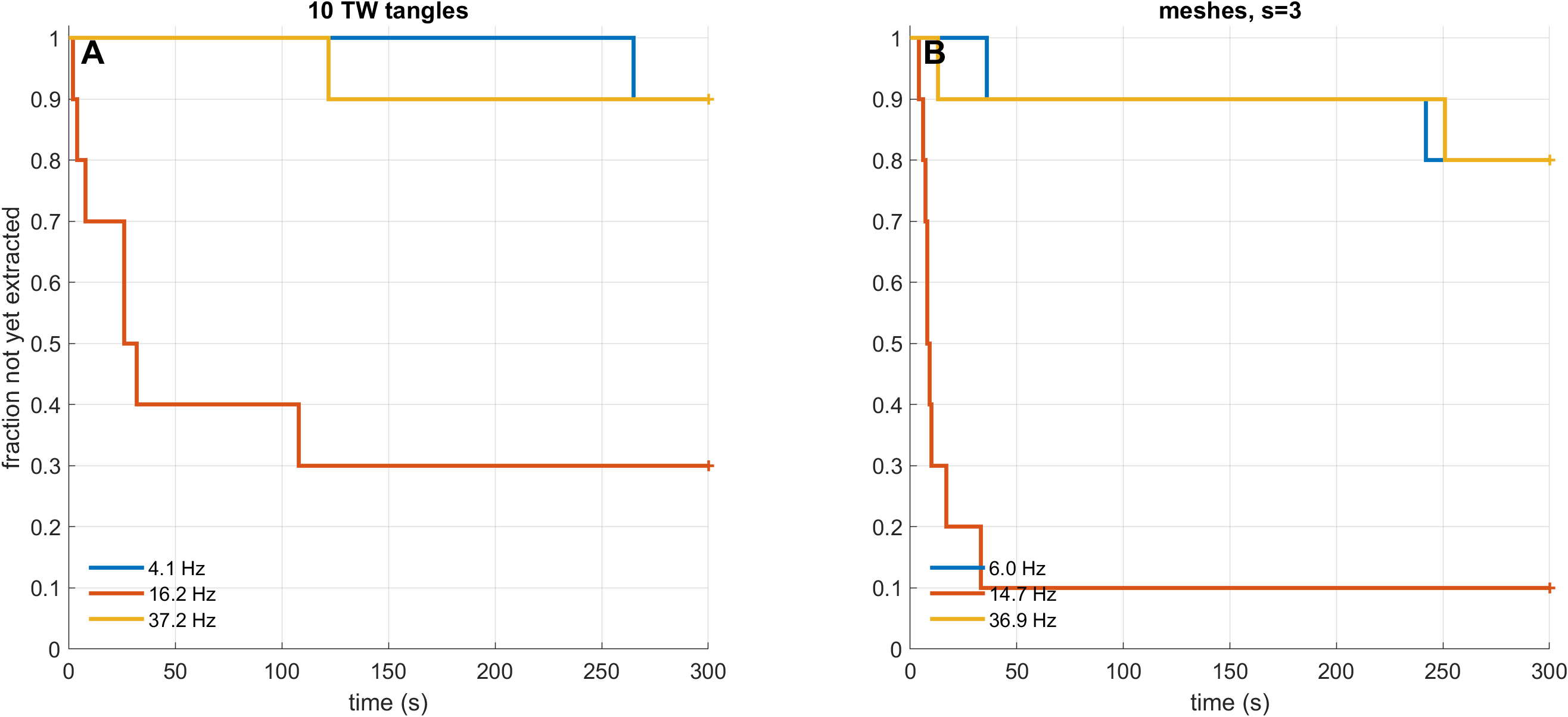}
    \caption{Time to extraction for representative drive levels. \textbf{A} 10-TW stochastic tangles. \textbf{B} Deterministic meshes with $s=3$. Curves are Kaplan--Meier estimates with $n=10$ trials per drive level; plus symbols denote right censoring at 300~s. Frequencies in the legends are measured medians.}
    \label{fig:survival}
\end{figure}

The binary endpoint in the main text retains the complete set of $n=10$ trials at each condition. To show the extraction time information without forcibly assigning 300 seconds extraction time to samples that remained attached, Fig.~\ref{fig:survival} plots Kaplan--Meier estimates for representative low, intermediate, and high drive levels. A plus symbol marks a trial truncated at the 300-s observation limit. The intermediate-frequency curves fall earliest for both the 10-TW tangles and the $s=3$ meshes, consistent with the probability maximum while preserving the finite observation window.

\subsection{Three-dimensional tracking of tangle center-of-mass motion} \label{sec:3d_tracking}

\begin{figure}[h]
    \centering
    \includegraphics[width=0.8\linewidth]{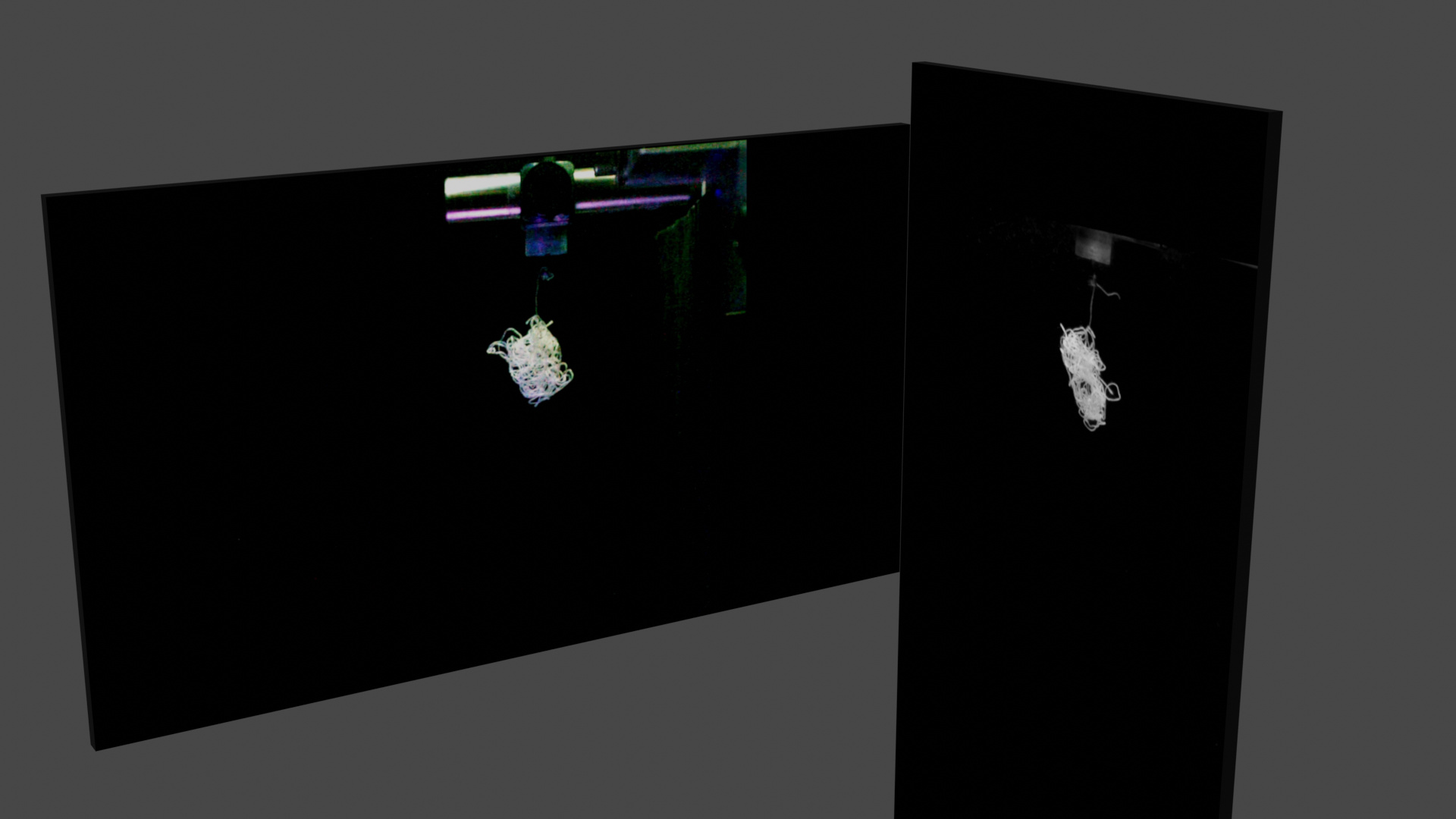}
    \caption{Representative visualization of the two orthogonal camera views used for three-dimensional reconstruction. The two image planes are shown perpendicular to one another to illustrate how the projected coordinates $(x,y)$ and $(z,y)$ were obtained. The piston diameter was used as the calibration length scale in both views.}
    \label{fig:3d_tracking_setup}
\end{figure}

To quantify the spatial excursions of the driven tangles, we reconstructed exemplar three-dimensional center-of-mass trajectories from synchronized high-speed videos. The tangle was recorded simultaneously from two approximately orthogonal viewing directions using two Chronos high-speed cameras operating at $250$ frames per second. One camera view was used to measure the projected coordinates $(x,y)$ of the tangle center of mass, whereas the second orthogonal view was used to measure $(z,y)$. A representative visualization of the two orthogonal views is shown in Fig.~\ref{fig:3d_tracking_setup}.

The two cameras were synchronized experimentally using a light pulse. Recording was first started on both cameras, and the light was then switched on. The frame at which the light intensity changed was identified in both recordings and used as a common time marker. This allowed the two videos to be aligned frame by frame before reconstructing the three-dimensional trajectory.

For each camera view, the image sequence was processed to isolate the visible tangle from the background and from the moving piston. A region of interest was selected around the suspended tangle, and each frame was converted into a binary mask corresponding to the tangle body. The projected center of mass in each view was then computed from the centroid of the segmented tangle pixels, yielding time series for the projected coordinates in the two camera planes.

Pixel coordinates were converted to physical units using the diameter of the cylindrical piston as the calibration length scale. Because the piston was cylindrical, its apparent diameter remained the same in both orthogonal views, providing a common calibration for the two recordings.

The final three-dimensional trajectory was obtained by combining the two synchronized projected trajectories. The front view provided $x(t)$ and $y(t)$, while the orthogonal view provided $z(t)$ and an independent measure of $y(t)$. The duplicate measurement of the vertical coordinate served as a consistency check on the synchronization and segmentation. The resulting trajectories were used to visualize the low-, intermediate- and high-frequency response regimes reported in the main text.

\subsection{High-speed measurement of disturbance propagation}
\label{sec:si_wave_propagation}

We analyzed six high-frequency recordings acquired at $10{,}000$ frames per second: three independently prepared 2-TW tangles and three independently prepared 10-TW tangles. All six recordings use $8.65\,$pixels/mm, calibrated from the piston diameter of 12.72~mm.
Dense Farneb\"ack optical flow is computed between successive grayscale frames, projected onto the radial unit vector from the tether-side origin, and multiplied by the frame rate. We analyze semicircular annuli on the tangle side of this origin. Positive radial velocity denotes motion away from the origin and negative velocity denotes motion toward it. The field is causally smoothed with an exponential weight $\alpha=1/3$ and averaged into annuli of width 3~pixels (0.347~mm). Annuli are retained only if they contain at least 20 pixels and cover at least 60\% of the nominal semicircular annulus. 
The speed estimator identifies eight-connected components of inward motion, $v_r\leq-5\,{\rm mm\,s^{-1}}$, in the valid radial bins. Components must span at least 20~mm radially. Within each component, one centerline time is the median time of its pixels at each radial row. The inner time $t_{\rm in}$ is the median of these centerline times within 0.5~mm of the component's minimum radius; $t_{\rm out}$ is defined analogously within 0.5~mm of its maximum radius. The event speed is
\begin{equation}
    v_{\rm event}=
    \frac{r_{\max}-r_{\min}}{t_{\rm out}-t_{\rm in}},
    \qquad t_{\rm out}>t_{\rm in}.
    \label{eq:si_endpoint_speed}
\end{equation}
The particle motion is inward, toward the tether-side origin, while the timing of this response progresses outward to larger radius. We retain components satisfying the valid-bin mask, inward-velocity threshold, minimum radial span and positive endpoint time order, with a finite computed speed.

\begin{figure}[t]
    \centering
    \includegraphics[width=\linewidth]{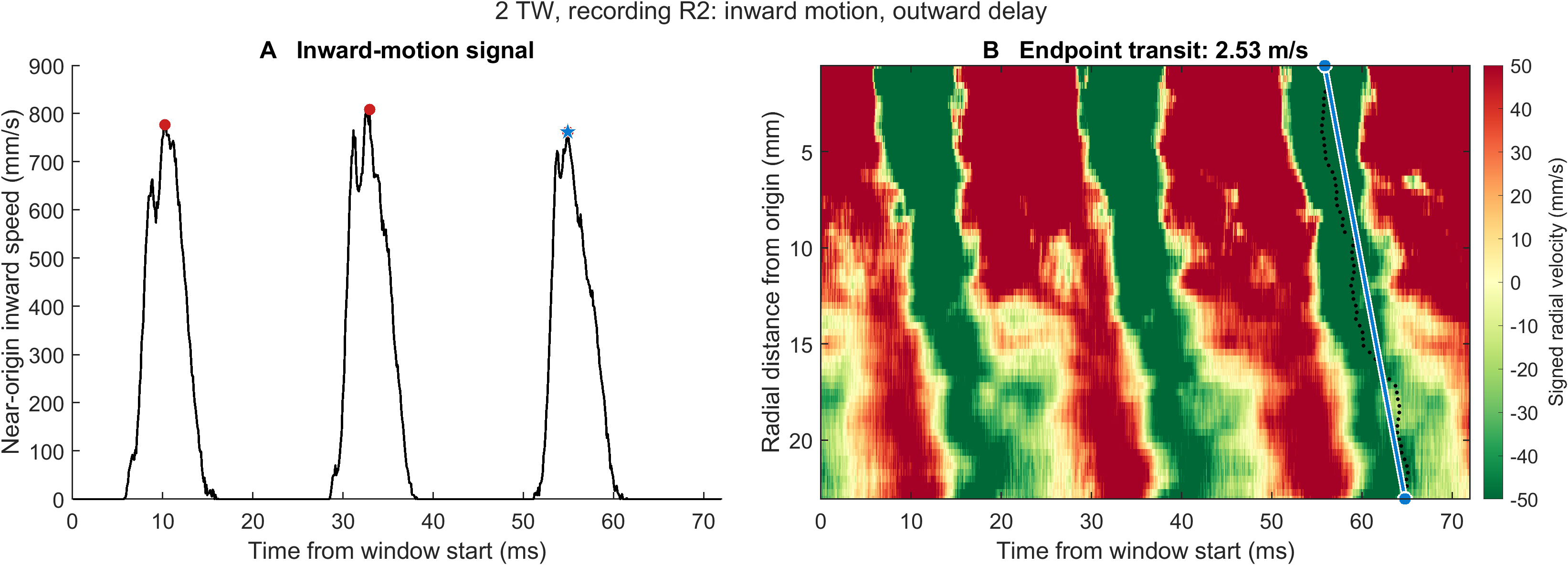}
    \caption{\textbf{Endpoint-transit measurement of an inward-motion disturbance in 2-TW recording R2.} \textbf{A} Mean inward-speed magnitude over valid $r=1$--$3\,$mm bins during the displayed window. Red markers identify signal peaks; the blue star identifies the peak nearest the component shown in B. \textbf{B} Signed radial-velocity kymograph over the same three events, with an inverted y-axis for radius. Black points show the median-time centerline of one component. Blue endpoints and their connecting segment give its propagation speed of 2.5~m/s using Eq.~\eqref{eq:si_endpoint_speed}.}
    \label{fig:wave_speed_ensemble}
\end{figure}

\begin{table}[h]
\centering
\caption{Endpoint-speed summaries. Events are repeated components nested within independent recordings.}
\begin{tabular}{lrrrr}
\hline
Recording & accepted events & median (m/s) & Q25 (m/s) & Q75 (m/s)\\
\hline
2 TW R1 & 75 & 3.2 & 2.4 & 4.8\\
2 TW R2 & 97 & 2.5 & 2.4 & 2.6\\
2 TW R3 & 113 & 2.6 & 2.4 & 3.6\\
10 TW R1 & 79 & 4.1 & 3.5 & 4.6\\
10 TW R2 & 87 & 3.4 & 2.9 & 3.7\\
10 TW R3 & 107 & 3.5 & 2.9 & 4.6\\
\hline
\end{tabular}
\label{tab:si_endpoint_summary}
\end{table}
We get six recording medians, one per independently prepared tangle, for the hierarchical summary in main-text Fig.~3D. The 2-TW condition is $2.8\pm0.4\,{\rm m\,s^{-1}}$ and the 10-TW condition is $3.6\pm0.4\,{\rm m\,s^{-1}}$ (mean $\pm$ sample standard deviation of three recording medians). Equal weighting of the two condition means gives $v_{\rm eff}=3.2\,{\rm m\,s^{-1}}$; the sample standard deviation of all six recording medians is $0.6\,{\rm m\,s^{-1}}$. Recording-level whiskers span within-recording event IQRs, whereas condition and overall whiskers show between-recording standard deviations.\\ For the main-text schematic, the phase delay is
\begin{equation}
    \phi=\frac{2\pi fL}{v_{\rm eff}},
    \qquad \frac{\phi}{2\pi}=\frac{fL}{v_{\rm eff}}
    \quad\hbox{drive cycles}.
    \label{eq:si_phase_delay}
\end{equation}
With $L=27\,$mm and $v_{\rm eff}=3.2\,{\rm m\,s^{-1}}$, the quarter- and half-cycle delays give $f_{\pi/2}=29.8\,$Hz and $f_\pi=59.5\,$Hz, delimiting the transmission crossover window.

\subsection{Architectural controls of the high-frequency response}
\label{sec:si_architectural_controls}

We used two qualitative controls to test how the transmission distance and rigidity of the supporting body affect the high-frequency response. In the compacted-tangle control, the same filament mass used for an extended tangle was compressed into a tight body while retaining the soft tethering thread. This "\emph{point-mass}" geometry substantially shortened the distance over which motion initiated at the tether had to spread for the body to move together. Independent repeats showed sustained agitation at representative low, intermediate and high drive frequencies; a representative sequence is shown in Supplementary Video~4.

In the rigid-mesh control, the compliant gauze body was replaced by a rigid plastic mesh with mass comparable to the tangle, while the thread stitched through the mesh remained soft. The qualitative response was reproduced in independent data sets and recorded across the tested drive frequencies. The rigid supporting body remained strongly agitated at high frequency, as shown in Supplementary Video~5. 

We also compared an extended stitched mesh with configurations whose wings were folded in half to reduce its extent at a fixed intermediate drive frequency. The extended mesh exhibited internal deformation, while the folded configurations developed center-of-mass jumps (Supplementary Video~6). These controls are useful in creating and testing the transmission picture of the tangle's response. 

\section{Lumped-parameter model for driven tangle dynamics}\label{sec:si_model}
To interpret the frequency-dependent response of the suspended tangles, we used a minimal lumped-parameter model. The tangle is represented as a point mass $m$ connected to the vertically driven suspension point by an effective filamentary spring of rest length $L_0$ and stiffness $k$. The model does not resolve the individual threads or their contact network. Instead, it coarse-grains the tangle into an effective mass, a tension-bearing element and dissipative terms that represent drag and internal losses.

The suspension point is driven vertically with displacement
\begin{equation}
    y_s(t)=A\sin(\omega t),
\end{equation}
where $A$ is the driving amplitude and $\omega$ is the imposed angular frequency. The position of the tangle center of mass is denoted by
\begin{equation}
    \mathbf{x}(t)=(x(t),y(t),z(t)).
\end{equation}
The vector from the driven suspension point to the mass is
\begin{equation}
    \mathbf{r}=(x,y-y_s,z),
    \qquad
    r=|\mathbf{r}|,
    \qquad
    \hat{\mathbf{r}}=\mathbf{r}/r.
\end{equation}
The effective spring is tension-only. It can pull on the mass when the distance from the suspension point exceeds the rest length, but it cannot push when the tether becomes slack. The spring force is therefore
\begin{equation}
    \mathbf{F}_s=-k(r-L_0)\hat{\mathbf{r}},
    \qquad r>L_0,
\end{equation}
and $\mathbf{F}_s=0$ for $r\leq L_0$. This unilateral spring law captures the main mechanical constraint of the experiment: the tethered thread can transmit tensile loads to the tangle, but it cannot transmit compressive loads.

We include two dissipative terms. The first is an isotropic
velocity-proportional drag acting on the mass at all times. The second acts
along the instantaneous connector direction and represents unresolved
losses associated with stretching, relaxation and sliding of fibers within
the tangle. We define the relative velocity between the mass and the driven
suspension point as
\begin{equation}
    \mathbf{v}_{\rm rel}
    =
    (\dot{x},\dot{y}-\dot{y}_s,\dot{z}),
\end{equation}
and its radial component as
\begin{equation}
    v_r=\mathbf{v}_{\rm rel}\cdot\hat{\mathbf{r}}.
\end{equation}

The elastic and radial-damping contributions together define a net
tensile force. Because the connector cannot transmit compression, the
transmitted tension is
\begin{equation}
    \mathcal{T}(r,v_r)
    =
    \begin{cases}
    \max\!\left\{
        0,\,
        k(r-L_0)+c_{\rm rad}v_r
    \right\},
    & r>L_0,\\[4pt]
    0,
    & r\leq L_0.
    \end{cases}
    \label{eq:dimensional_unilateral_tension}
\end{equation}
The equation of motion is therefore
\begin{equation}
    m\ddot{\mathbf{x}}
    =
    -\mathcal{T}(r,v_r)\hat{\mathbf r}
    -c\dot{\mathbf{x}}
    -mg\hat{\mathbf{y}}
    +m\mathbf{a}_{\rm noise},
    \label{eq:lumped_parameter_model_eom}
\end{equation}
where $c$ is the isotropic damping coefficient, $c_{\rm rad}$ is the radial
damping coefficient and $\mathbf{a}_{\rm noise}$ is a fluctuating
acceleration. During inward radial motion, $v_r<0$, radial damping reduces
the transmitted tension. If the damping contribution exceeds the elastic
contribution, the connector unloads rather than exerting a compressive
force.

We nondimensionalize length by $L_0$ and time by the spring--mass time scale
\begin{equation}
    T^*=\sqrt{\frac{m}{k}}.
\end{equation}
The dimensionless variables are $\mathbf{X}=\mathbf{x}/L_0$ and $\tau=t/T^*$. The driven suspension point becomes
\begin{equation}
    Y_s(\tau)=\epsilon\sin(\Omega^*\tau),
    \qquad
    \epsilon=\frac{A}{L_0},
    \qquad
    \Omega^*=\omega T^*.
\end{equation}
We use the stiffness-to-weight ratio
\begin{equation}
    \chi=\frac{kL_0}{mg},
\end{equation}

so that gravity enters as $\chi^{-1}$. The dimensionless damping
coefficients are
\begin{equation}
    \zeta=\frac{cT^*}{m},
    \qquad
    \zeta_{\rm rad}=\frac{c_{\rm rad}T^*}{m}.
\end{equation}
With
\begin{equation}
    \mathbf{R}=(X,Y-Y_s,Z),
    \qquad
    D=|\mathbf{R}|,
    \qquad
    \hat{\mathbf{R}}=\mathbf{R}/D,
\end{equation}
the dimensionless velocity and relative radial velocity are
\begin{equation}
    \mathbf{U}=\frac{d\mathbf{X}}{d\tau},
    \qquad
    V_r=(U_X,U_Y-\dot{Y}_s,U_Z)\cdot\hat{\mathbf{R}}.
\end{equation}
The dimensionless transmitted tension is
\begin{equation}
    \mathcal{T}^*(D,V_r)
    =
    \begin{cases}
    \max\!\left\{
        0,\,
        (D-1)+\zeta_{\rm rad}V_r
    \right\},
    & D>1,\\[4pt]
    0,
    & D\leq1.
    \end{cases}
    \label{eq:dimensionless_unilateral_tension}
\end{equation}
The dimensionless equation of motion is
\begin{equation}
    \frac{d^2\mathbf{X}}{d\tau^2}
    =
    -\mathcal{T}^*(D,V_r)\hat{\mathbf{R}}
    -\zeta\mathbf{U}
    -\chi^{-1}\hat{\mathbf{Y}}
    +\boldsymbol{\eta}(\tau).
    \label{eq:dimensionless_eom}
\end{equation}

The three drive coordinates obey the exact identity
\begin{equation}
    \Gamma_{\rm acc}=\frac{A\omega^2}{g}
    =\epsilon\chi(\Omega^*)^2.
    \label{eq:si_coordinate_identity}
\end{equation}
Thus $\Gamma_{\rm acc}$ and $\chi$ are independent model controls when $\epsilon$ is held fixed, while $\Omega^*$ is derived. At the fixed experimental amplitude, $f$ and $\Gamma_{\rm acc}$ are one-to-one; an amplitude sweep would be required to test complete experimental collapse at fixed acceleration.

We add a small correlated acceleration noise to break perfect symmetry and
represent unresolved fluctuations. It is modeled as an
Ornstein--Uhlenbeck process,
\begin{equation}
    d\boldsymbol{\xi}
    =
    -\frac{1}{\tau_c}\boldsymbol{\xi}\,d\tau
    +
    \sqrt{\frac{2}{\tau_c}}\,d\mathbf{W},
    \qquad
    \boldsymbol{\eta}(\tau)
    =
    \sigma_a\boldsymbol{\xi}(\tau).
\end{equation}
The process was initialized in its stationary distribution,
$\boldsymbol{\xi}(0)\sim\mathcal{N}(\mathbf{0},\mathbf{I})$.

We chose the experimental value of the drive amplitude as our physical drive amplitude in the model while the reference mass of a tangle and typical connector length set the dimensional scale. The quasi-static loading slopes located the experimental markers onto the phase diagram in main-text Fig. 4. The damping and noise values below were chosen as follows. For the response map in main-text Fig.~4, the parameters were
$\epsilon=0.15$, $\zeta_{\rm rad}=0.12$,
$\sigma_a=5\times10^{-4}$ and $\tau_c=0.05$. The noise amplitude and
correlation time were held fixed in dimensionless elastic-time units
throughout the parameter scan. Thus, the stochastic forcing represents a
uniform perturbation of the dimensionless dynamics rather than a fixed
dimensional acceleration.

The isotropic damping followed
$\zeta=c_{\rm air}/\sqrt{mk}$, with
$c_{\rm air}=1.8\times10^{-5}\,{\rm kg\,s^{-1}}$, and therefore varied with
$\chi$. The reference dimensional values were
$m=5\times10^{-5}\,{\rm kg}$, $L_0=0.05\,{\rm m}$ and
$g=9.8\,{\rm m\,s^{-2}}$, giving
$A=\epsilon L_0=0.75\,{\rm cm}$.

The simulations were run directly on a rectangular grid of 80 logarithmically spaced values of $\Gamma_{\rm acc}$ from 0.2 to 160 and 40 logarithmically spaced values of $\chi$ from 10 to 1000. At each requested pair we set
\begin{equation}
    k=\frac{\chi mg}{L_0},\qquad
    \Omega^*=\sqrt{\frac{\Gamma_{\rm acc}}{\epsilon\chi}},
\end{equation}
and integrated the elastic-time equation above. The plotted field is therefore native to the independent $(\Gamma_{\rm acc},\chi)$ grid rather than an interpolated transformation of a scan in $\Omega^*$. For the reference dimensions, the acceleration range corresponds to $f=2.57$--$72.8\,$Hz, the stiffness range to $k=0.098$--$9.8\,{\rm N\,m^{-1}}$, and the derived normalized frequency spans $\Omega^*=0.0365$--$10.33$ across the map.

Each realization was integrated for $10\,{\rm s}$ of physical time from
$\mathbf{X}(0)=(0,-1,0)$ and $\mathbf{U}(0)=(0,0,0)$ with a dimensionless time step $d\tau=10^{-3}$. 

Ten independent noise realizations were simulated at every point in
parameter space, with an independent pseudorandom substream assigned to
each parameter--realization pair. For a step of duration $h=d\tau$, the
Ornstein--Uhlenbeck state was advanced according to
\begin{equation}
    \boldsymbol{\xi}_{n+1}
    =
    \alpha\boldsymbol{\xi}_{n}
    +
    \sqrt{1-\alpha^2}\,\mathbf{G}_n,
    \qquad
    \alpha=\exp(-h/\tau_c),
\end{equation}
where $\mathbf{G}_n\sim\mathcal{N}(\mathbf{0},\mathbf{I})$. The acceleration
used during the mechanical step was
\begin{equation}
    \boldsymbol{\eta}_n
    =
    \frac{\sigma_a}{2}
    \left(
        \boldsymbol{\xi}_n+\boldsymbol{\xi}_{n+1}
    \right),
\end{equation}
and was held constant over each mechanical time step. All realizations
were simulated for the same physical observation time.

The model does not explicitly describe filament slip or extraction.
Instead, its response is quantified by the excursion of the suspended mass
about its temporal mean. This quantity characterizes the spatial extent of
the driven motion and is not identified with an extraction probability.
For realization $q$, let
$\mathbf{X}_q(\tau)=[X_q(\tau),Y_q(\tau),Z_q(\tau)]$ denote the
dimensionless center-of-mass trajectory. Its RMS excursion is
\begin{equation}
    \frac{A_{{\rm rms},q}}{L_0}
    =\left(
    \sigma_{X,q}^2+\sigma_{Y,q}^2+\sigma_{Z,q}^2
    \right)^{1/2},
    \label{eq:si_rms_excursion}
\end{equation}
where each $\sigma$ is the temporal standard deviation over the 10-s trajectory. The coordinate components are combined within each realization before averaging,
\begin{equation}
    \frac{\langle A_{\rm rms}\rangle}{L_0}
    =\frac{1}{10}\sum_{q=1}^{10}
    \left(
    \sigma_{X,q}^2+\sigma_{Y,q}^2+\sigma_{Z,q}^2
    \right)^{1/2}.
    \label{eq:si_mean_rms_excursion}
\end{equation}
Main-text Fig.~4B displays the dimensionless ensemble-averaged RMS
excursion $\langle A_{\rm rms}\rangle/L_0$. Small values indicate localized motion, whereas large
values indicate broad trajectories associated with repeated slackening and
recapture. 

The representative trajectories in main-text Fig.~4C--E were evaluated at
a common stiffness-to-weight ratio $\chi=200$. The low-, intermediate- and
high-frequency examples correspond respectively to
\begin{equation}
\begin{aligned}
    (\Gamma_{\rm acc},\chi)_{\rm low}
    &=(0.30,200),\\
    (\Gamma_{\rm acc},\chi)_{\rm intermediate}
    &=(7.97,200),\\
    (\Gamma_{\rm acc},\chi)_{\rm high}
    &=(150,200).
\end{aligned}
\end{equation}
Trajectory color represents the speed of the mass relative to the moving
support,
\begin{equation}
    v_{\rm rel}
    =
    \left|
        \dot{\mathbf{x}}
        -
        \dot{y}_s\hat{\mathbf{y}}
    \right|.
\end{equation}
Supplementary Video~7 shows the corresponding transition from continuous
loading to intermittent slack--recapture and high-frequency localization.

The experimental ordinates use median quasi-static loading slopes measured from the rising portions of the force--displacement curves. The values used in the plotting script are
\begin{center}
\begin{tabular}{lcc}
\hline\hline
Condition & $k_{\rm eff}$ (N/m) & $\chi=k_{\rm eff}L_0/(mg)$ \\
\hline
0 TW & 1.8 & 184 \\
2 TW & 4.0 & 408 \\
10 TW & 5.0 & 510 \\
$s=2$ mesh & 5.5 & 561 \\
$s=3$ mesh & 7.4 & 755 \\
\hline\hline
\end{tabular}
\end{center}
The abscissae are calculated independently from the experimental frequencies using $\Gamma_{\rm acc}=A(2\pi f)^2/g$. The blue propagation window in main-text Fig.~4B uses the tangle reference propagation speed and the projected average $L=27\,$mm for all $\chi$, with the band limits set by the $\pi/2$ and $\pi$ phase delays. 

\section{Model-based parameter inference from experimental trajectories}\label{sec:si_inference}

\begin{figure}[h]
    \centering
    \includegraphics[width=\linewidth]{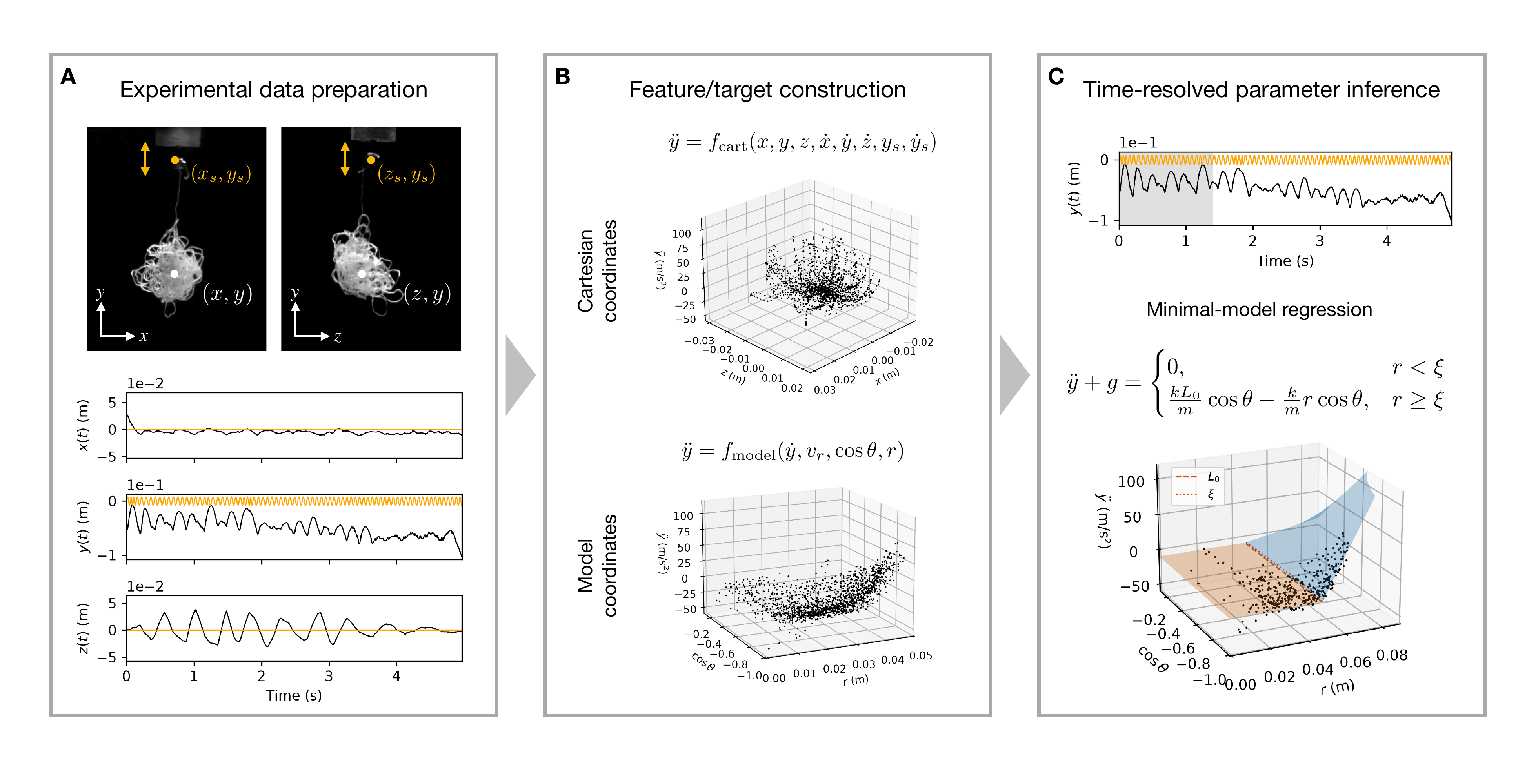}
    \caption{Model-based parameter inference pipeline. \textbf{A.} Experimental images of the tangle under orthogonal camera views (top). Time series prepared from 3D tracking and Gaussian-kernel smoothing for a 10 TW sample at moderate-frequency driving, with selected bandwidths $h_x^*=6.81$ ms, $h_y^*=4.64$ ms, and $h_z^*=5.62$ ms (bottom). \textbf{B.} Scatter plots of the measured vertical accelerations $\ddot{y}$ against the Cartesian coordinates $(x, z)$ and the transformed coordinates $(r, \cos\theta)$. \textbf{C.} Highlighted time window in the time series (top) and the corresponding fitted-surface plot for the vertical acceleration $\ddot{y}$ as a function of the transformed coordinates $(r, \cos\theta)$, with data points overlaid as scatter points (bottom). The inferred rest length $L_0$ and the breakpoint $\xi$ are shown as dashed and dotted lines, respectively. For visualization, the fitted surface is shown in blue for taut observations ($r>L_0$) and orange for slack observations ($r<L_0$). The regression model shown in this schematic is the tension-only elastic model in Eq.~(\ref{eq:regression-elastic}); the full regression model in Eq.~(\ref{eq:regression-full}) also includes isotropic and radial damping contributions.}
    \label{fig:inference_pipeline}
\end{figure}

The minimal model in the previous section provides a reduced description of the tangle dynamics that can be studied through simulations. We also fit its tension-only equation directly to experimentally measured center-of-mass trajectories, using the result as a time-resolved phenomenological description of the driven state. In this section, we present an analysis pipeline that performs this model-based parameter inference (Fig.~\ref{fig:inference_pipeline}). The pipeline preprocesses the 3D trajectories, transforms the motion into coordinates relative to the oscillating piston, and fits the equation of motion over successive time windows. To prevent these regression coefficients from being confused with the quasi-static loading stiffness, we denote them by $\kappa_{\rm dyn}(t)\equiv k/m$ and $\ell_{\rm dyn}(t)\equiv L_0$. They describe local fits to the observed motion and are not used to place experimental points on the main-text response map.

\subsection{Experimental trajectory preparation} \label{sec:exp_prep}

We begin with the reconstructed 3D center-of-mass trajectories described in Sec.~\ref{sec:3d_tracking} and depicted in Fig.~\ref{fig:inference_pipeline}A. These trajectories provide the measured tangle position $\mathbf{x}(t)=(x(t),y(t),z(t))$ for each driving experiment, and the imposed vertical motion of the suspension point, $y_s(t)$, was obtained from the recorded piston signal and calibrated using the measured piston displacement amplitude. The tangle center-of-mass trajectories together with the imposed suspension-point motion form the input data to the pipeline.

To prepare the experimental data for parameter inference, we perform a smoothing step to reduce noise in the measured trajectories. The minimal model is expressed as an equation of motion in Eq.~(\ref{eq:lumped_parameter_model_eom}), and fitting it to experimental data requires estimating time derivatives of the measured center-of-mass trajectory. This makes smoothing necessary as a processing step, because numerical differentiation amplifies measurement noise. For example, if a measured coordinate is written as $q_i=q(t_i)+\eta_i$, where $\eta_i$ is the measurement noise, then a finite-difference first derivative contains terms of order $(\eta_{i+1}-\eta_i)/\Delta t$, while a second derivative contains terms of order $(\eta_{i+1}-2\eta_i+\eta_{i-1})/\Delta t^2$. Thus, even small frame-to-frame fluctuations in the trajectories can produce much larger errors in the estimated velocities and accelerations. We therefore first reconstruct smooth coordinate trajectories from the measured positions before computing the derivatives used in the inference procedure. 

In the analysis pipeline, this smoothing step was performed using a local Gaussian-kernel smoother applied independently to each measured coordinate of the center-of-mass trajectory. For a coordinate $q(t)$ sampled at times $t_i$, the smoothed value at time $t$ was computed as a weighted average of nearby measurements,
\begin{equation}
\label{eq:kernel-smoother}
\hat q_h(t)=\frac{\sum_i K_h(t-t_i)q_i}{\sum_i K_h(t-t_i)},  
\end{equation}
where the Gaussian kernel is $K_h(t-t_i)=\exp\left(-(t-t_i)^2/2h^2\right)$.
The bandwidth $h$ controls the temporal smoothing scale; a smaller bandwidth uses only nearby frames and follows the measured trajectory more closely, while a larger bandwidth averages over a wider time window and produces a smoother trajectory. 

We use cross-validation to select the bandwidth $h$ for each coordinate $q\in\{x,y,z\}$. For each candidate bandwidth $h$ in the set of candidate bandwidths $\mathcal H$, the trajectory samples are split into training and validation subsets. For split $s = 1,\dots,S$, the smoother $\hat{q}_h^{(s)}(t)$ is fit using the training samples and evaluated by the mean squared error on the held-out samples:
\begin{equation}
    \mathrm{MSE}_s(h)
    =
    \frac{1}{|V_s|}
    \sum_{i\in V_s}
    \left[q_i-\hat q_h^{(s)}(t_i)\right]^2,
\end{equation}
where $V_s$ is the set of validation indices for split $s$. The bandwidth with the lowest average validation error across the cross-validation splits is selected:
\begin{equation}
    h^*_q = \arg\min_{h \in \mathcal{H}} \frac{1}{S} \sum_{s=1}^S \mathrm{MSE}_s(h).
\end{equation}
The smoother is then refit using all measured samples for that coordinate with bandwidth $h_q^*$ to reconstruct the final smoothed trajectory. 

A representative example of the smoothed 3D trajectories is shown in Fig.~\ref{fig:inference_pipeline}A, and the selected bandwidths for the 10 TW trajectories used for the analysis in Fig.~\ref{fig:inference_results} are listed in Table~\ref{tab:smoothing_bandwidths}. The Gaussian-kernel smoother is a local method that only assumes that the underlying positions vary smoothly over short time windows. This is well suited to the experimental trajectories because the tangle motion is irregular and can change over the course of a trial, while a global sinusoidal or Fourier representation would impose a stronger functional assumption on the trajectory. The bandwidth also has a direct interpretation as a smoothing time scale, and cross-validation provides a reproducible way to select this scale from the data. As the time series in Fig.~\ref{fig:inference_pipeline}A show, this smoothing and bandwidth selection procedure reduces frame-to-frame noise while preserving local trajectory features that are important for characterizing the effective response.

\begin{table}[h]
    \centering
    \caption{Selected Gaussian-kernel bandwidths for the 10 TW trajectories used in Fig.~\ref{fig:inference_results}.}
    \label{tab:smoothing_bandwidths}
    \begin{tabular}{c c c c}
        \hline
        Driving regime & $h_x^*$ (ms) & $h_y^*$ (ms) & $h_z^*$ (ms) \\
        \hline
        Low      & 5.62 & 4.64 & 5.62 \\
        Moderate & 6.81 & 4.64 & 5.62 \\
        High     & 5.62 & 3.16 & 6.81 \\
        \hline
    \end{tabular}
\end{table}

\subsection{Coordinate transform and feature/target construction} \label{sec:feature_target}

After smoothing the measured trajectories, we transform the motion into the relative coordinates of the minimal model defined in Eq.~(\ref{eq:lumped_parameter_model_eom}). This step rewrites the center-of-mass motion in variables measured relative to the driven suspension point, which is the natural frame for the equation of motion. For instance, the vertical acceleration becomes a function of the model coordinates instead of the raw Cartesian coordinates,
\begin{equation}
\label{eq:relative-transform}
\ddot y = f_{\mathrm{cart}}(x,y,z,\dot x,\dot y,\dot z,y_s,\dot y_s)  
\quad \longrightarrow \quad  
\ddot y = f_{\mathrm{model}}(\dot y,v_r,\cos\theta,r),  
\end{equation}
where $\cos\theta=\hat{\mathbf{r}}\cdot\hat{\mathbf{y}}$ is the vertical projection of the instantaneous spring direction. Fig.~\ref{fig:inference_pipeline}B visualizes this coordinate transformation for a representative trajectory time window. When the measured vertical accelerations $\ddot{y}$ are plotted against the transformed coordinates $(r,\cos\theta)$, rather than the raw Cartesian coordinates $(x,z)$, the point cloud becomes structured in a way that suggests a more clearly defined dependence on the transformed variables.

In this representation, the parameter inference can be formulated as a regression problem whose fitted coefficients are directly tied to the effective parameters of the minimal model. Projecting Eq.~(\ref{eq:lumped_parameter_model_eom}) onto the vertical direction gives
\begin{equation}
\label{eq:vertical-eom}
\ddot y =
-g  
-\frac{c}{m}\dot y  
-\frac{c_{\mathrm{rad}}}{m}v_r\cos\theta  
+\mathbf{1}_{r>L_0}\left(\frac{kL_0}{m}\cos\theta-\frac{k}{m}r\cos\theta\right)  
+a_{\mathrm{noise},y}.  
\end{equation}
We focus on the vertical equation of motion because it contains all the force contributions from the minimal model. The regression target is the vertical acceleration $\ddot y(t)$, and the corresponding feature set contains the constant term, $\dot y$, $v_r\cos\theta$, $\cos\theta$, and $r\cos\theta$, along with $r$ for constructing the indicator function. These terms have a direct interpretation from the model: the constant term accounts for gravity, $\dot y$ for isotropic damping, $v_r\cos\theta$ for radial damping, $\cos\theta$ and $r\cos\theta$ give the two spring-force contributions, and $r$ itself contributes to the tension-only activation condition.

\subsection{Time-resolved parameter inference} \label{sec:param_inference}

The feature and target data constructed above are then used to estimate the effective parameters of the reduced model directly from the measured motion. We perform this inference separately over time windows, so that the fitted parameters can vary as the tangle evolves during a driving experiment (Fig.~\ref{fig:inference_pipeline}C). Within each window, the effective parameters are approximated as constants, while their variation from one window to the next tracks the evolution of the tangle dynamics over time.

For a fixed choice of taut--slack breakpoint $\xi$, the general regression equation in each time window can be written as
\begin{equation}
\label{eq:regression-full}
\ddot y_i =
-\beta_0
-\beta_1\dot y_i
-\beta_2 v_{r,i}\cos\theta_i
+\mathbf{1}_{r_i>\xi}\left(\beta_3\cos\theta_i-\beta_4 r_i\cos\theta_i\right)
+\varepsilon_i,
\end{equation}
where the subscripts denote observations in one time window and $\varepsilon_i$ collects unmodeled forces and derivative error. For a fixed breakpoint, the coefficients are estimated by ordinary least squares, giving $g=\beta_0$, $c/m=\beta_1$, $c_{\rm rad}/m=\beta_2$, $kL_0/m=\beta_3$, $k/m=\beta_4$, and $L_0=\beta_3/\beta_4$. The complete pipeline is not parameter-free: smoothing bandwidth, window length, and the cross-validated breakpoint affect the estimates. Because several force terms are correlated over a short trajectory, the fitted coefficients are effective local parameters and are not guaranteed to be uniquely identifiable physical constants.

For the parameter inference, we keep the taut--slack breakpoint $\xi$ separate from the inferred rest length $L_0$ (Fig.~\ref{fig:inference_pipeline}C). We select $\xi$ within each time window by cross-validation over a grid of candidate breakpoint values. The regression coefficients are fit for each candidate $\xi$, and the breakpoint with the lowest held-out prediction error for $\ddot y_i$ is used. This determines the taut--slack separation from the measured dynamics directly without manual selection.

The breakpoint $\xi$ determines which observations are treated as taut during the regression, while $L_0$ is obtained afterward from the fitted coefficients. If the breakpoint were forced to equal $L_0$, then the indicator function would depend on the same coefficients being estimated, making the regression nonlinear. This separation also reflects the fact that the measured center-of-mass motion is a coarse-grained description of the tangle, so the apparent onset of tension may not match the inferred rest length exactly. The two quantities can nevertheless be compared as a consistency check on the tension-only interpretation. In the representative example shown in Fig.~\ref{fig:inference_pipeline}C, the fitted breakpoint $\xi$ and the inferred rest length $L_0$ agree closely. 

After fitting the model in each time window, we estimate uncertainty ranges in the inferred parameters by bootstrap resampling. We also evaluate the quality of each fit using the mean squared error and the coefficient of determination $R^2$ for the predicted vertical acceleration. These provide complementary measures of the inference: parameter uncertainty from bootstrap resampling indicates the stability of the inferred parameters, while the fit-quality metrics indicate how well the minimal model captures the observed motion of the tangle.

\subsection{Time-dependent effective model parameters} \label{sec:time_dep_params}

\begin{figure}[t]
    \centering
    \includegraphics[width=\linewidth]{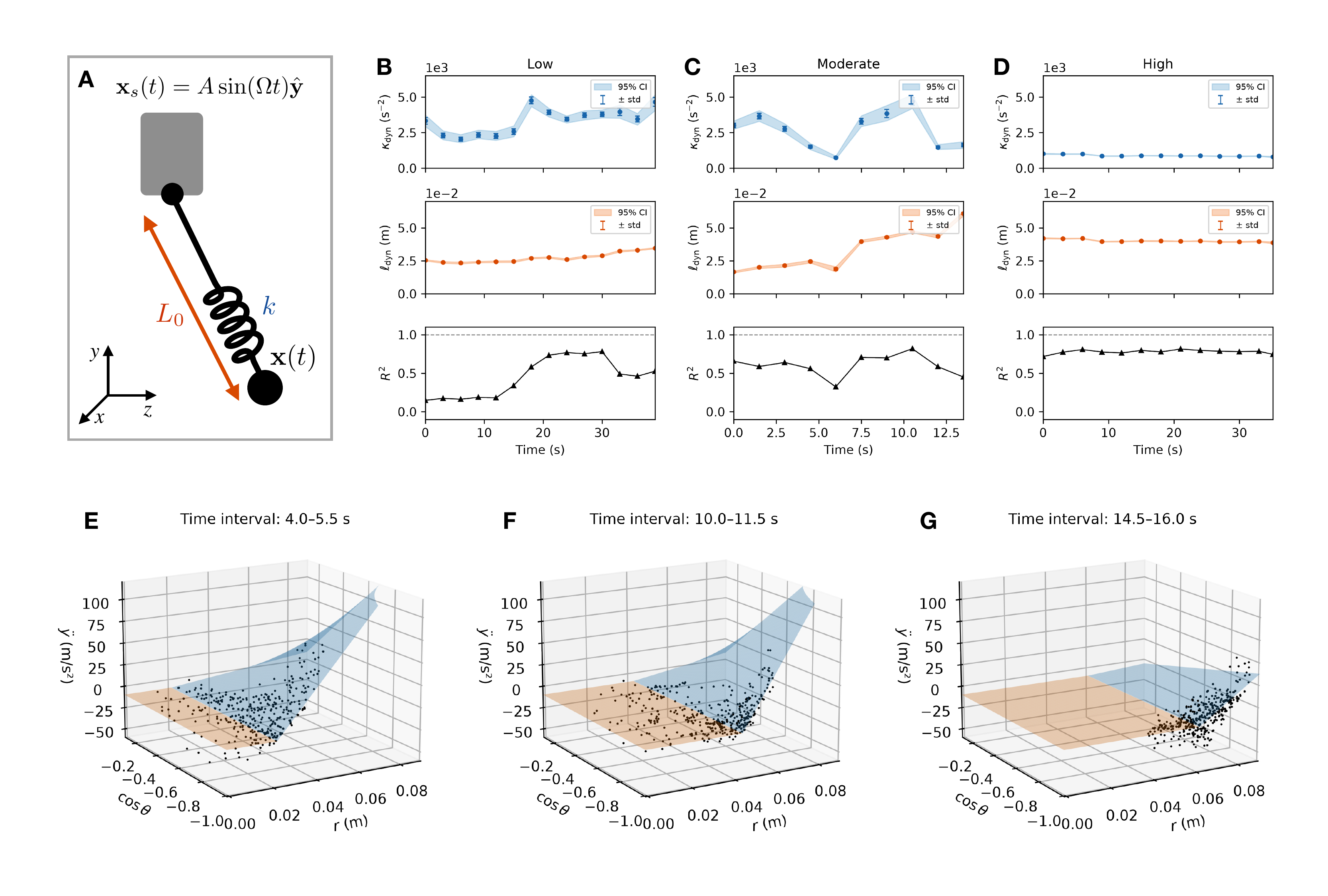}
    \caption{Time-resolved effective parameters inferred from experimental trajectories. \textbf{A.} Schematic of the minimal model restricted to the tension-only elastic forces, following the vertical equation of motion in Eq.~(\ref{eq:regression-elastic}). \textbf{B--D.} Inferred dynamic coefficient $\kappa_{\rm dyn}\equiv k/m$ (units $\mathrm{s^{-2}}$) and effective length $\ell_{\rm dyn}\equiv L_0$ for representative 10 TW trials driven at low, moderate, and high frequencies. Shaded regions indicate 95\% bootstrap confidence intervals, and pointwise error bars indicate one bootstrap standard deviation. \textbf{E--G.} Fitted-surface plots for selected time windows in the moderate-frequency trial in panel C.}
    \label{fig:inference_results}
\end{figure}

The analysis procedure and regression framework formulated in Sections \ref{sec:exp_prep}--\ref{sec:param_inference} allow all parameters in the minimal model to be estimated in principle, including contributions from isotropic damping, radial damping, and tension-only elastic forces. Here, we focus on results where the inference is restricted to the elastic contributions, and gravity is fixed at its physical value $g = 9.81\,\mathrm{m\,s^{-2}}$:
\begin{equation}
\label{eq:regression-elastic}
\ddot y_i + g =
\mathbf{1}_{r_i>\xi}\left(\beta_3\cos\theta_i-\beta_4 r_i\cos\theta_i\right)
+\varepsilon_i.
\end{equation}

A schematic of this model is shown in Fig.~\ref{fig:inference_results}A. This restriction reduces variance and focuses the inference on the dominant contributions to the observed motion coming from the tension-only elastic forces. The damping contributions in the experimental trajectories were smaller and more sensitive to noise in the velocity estimates, which made the fitted coefficients less stable across time windows. We therefore treat the remaining damping contributions as part of the error term, with fit quality assessed using the metrics discussed in Section \ref{sec:param_inference}.

The trajectory-inferred $\kappa_{\rm dyn}=k/m$ is a local phenomenological coefficient obtained after smoothing and differentiation. It is not the quasi-static loading slope: the two procedures sample different motions and solve different inverse problems. Only the quasi-static $k_{\rm eff}$ is used to calculate $\chi$ and place samples on the main-text map; $\kappa_{\rm dyn}$ is used solely to track changes within the selected driven trajectories.

In Figs.~\ref{fig:inference_results}B--D, we show the time-dependent effective parameters inferred from representative 10 TW trajectories driven at low, moderate, and high frequencies. These inferred parameters provide a complementary, model-based view of the center-of-mass dynamics shown in main-text Fig. 2, showing how the effective elastic response evolves or remains stable over the course of a trial, and how its behavior varies across different driving regimes. 

For the representative low-frequency trial (Fig.~\ref{fig:inference_results}B), the center of mass follows the support with small excursions and limited radial deformation, leaving relatively few observations that strongly constrain the dynamic coefficient. The high-frequency representative (Fig.~\ref{fig:inference_results}D) has similarly limited radial motion. The apparent stability of the fitted coefficients in these examples should therefore be read together with the bootstrap intervals and fit metrics, not as proof of a time-invariant material stiffness.

The representative intermediate-frequency trial (Fig.~\ref{fig:inference_results}C) shows larger excursions, larger fluctuations in $\kappa_{\rm dyn}$, and an increase in $\ell_{\rm dyn}$ before extraction. These time correlations are consistent with progressive slipping and rearrangement, but a single trajectory does not establish a population-level frequency dependence of stiffness.

In Figs.~\ref{fig:inference_results}E--G, we present fitted-surface plots for selected time windows in the 10 TW, moderate-frequency trial in Fig.~\ref{fig:inference_results}C, with data points from each time window overlaid as scatter points. While the parameter curves summarize each time window by $\kappa_{\rm dyn}$ and $\ell_{\rm dyn}$, the fitted-surface plots show how the data are organized in the transformed coordinates. Data points lying close to the fitted surface indicate that the vertical equation of motion in Eq.~(\ref{eq:regression-elastic}) captures the center-of-mass dynamics well in that window, providing a visual counterpart to the MSE and $R^2$ fit metrics.

The coefficients $\kappa_{\rm dyn}$ and $\ell_{\rm dyn}$ also have clear visual interpretations in these fitted-surface plots. The dynamic coefficient controls the steepness of the fitted surface in the taut regime (blue surface), and Figs.~\ref{fig:inference_results}E--G show fluctuations over the course of the trial, as also observed in Fig.~\ref{fig:inference_results}C. In these fitted-surface plots, the rest length sets the $r$-axis location of the visual transition between the taut and slack regimes, represented by the blue and orange surfaces, respectively. Across the three panels, this breakpoint shifts from smaller to larger values of $r$, matching the increase in $\ell_{\rm dyn}$ in Fig.~\ref{fig:inference_results}C as the thread becomes progressively extracted.

\subsection{Summary and discussion} \label{sec:inference_summary}

In summary, the pipeline provides time-resolved effective coefficients for representative trajectories and visualizes where a tension-only regression approximates the measured vertical acceleration. The intermediate-frequency example shows stronger coefficient evolution than the selected low- and high-frequency examples. The fit is a within-trajectory diagnostic rather than an independent stiffness measurement or a test of resonance.

This inference framework probes the effective properties of the tangle while it is in motion and being driven. The inferred quantities describe the coarse-grained response of the driven tangle, and their time dependence can reflect changes in the effective elastic behavior as contacts rearrange and the tethered thread slips through the network. Because the time-resolved parameter estimates are obtained from the motion itself, they provide access to the evolving response during the experimental filament-extraction dynamics, while remaining distinct from static pullout measurements that characterize force buildup before slip.

Furthermore, the inference framework provides useful and interpretable diagnostics for the reduced model. The fit metrics quantify how well the model captures the tangle's center-of-mass dynamics during filament extraction, the parameter uncertainties indicate how well constrained the inferred quantities are by the data, and the fitted-surface plots provide a direct way to visualize and interpret the results. The parameter-inference procedure we presented therefore serves a dual purpose: to infer effective parameters from experimental trajectories, and to assess when the minimal model provides a useful description of the driven tangle motion.

\section{Gravity, elastic--inertial and propagation boundaries}
\label{sec:si_dynamical_boundaries}

The independent-coordinate representation makes it possible to estimate three physical boundaries of the driven response. The first determines whether the tether can become slack. The second compares the forcing period with the lumped elastic time, and the third compares it with a measured spatial propagation time. These estimates also clarify the roles of the two stiffness measurements used in this work. The quasi-static loading slope $k_{\rm eff}$ locates each experimental condition on the main-text phase map because it is measured independently of the driven trajectory and characterizes force buildup before slip. The trajectory-inferred coefficient $\kappa_{\rm dyn}$ describes the local coupling observed during a driven trial and is used below only as an order-of-magnitude check on the propagation scale.

For the imposed motion $y_s=A\sin(\omega t)$, the tension required to keep a slowly varying tether taut is
\begin{equation}
    T(t)=m[g+\ddot y_s(t)].
\end{equation}
Its minimum over a forcing cycle is
\begin{equation}
    T_{\min}=\rm{max}\{0, mg(1-\Gamma_{\rm acc})\}.
\end{equation}
Because the filament cannot support compression, $T_{\min}=0$ gives the gravity-mediated slackening boundary at
\begin{equation}
    \boxed{\Gamma_{\rm acc,on}=1.}
    \label{eq:si_slack_boundary}
\end{equation}
For $A=0.75\,$cm, this corresponds to $f_{\rm on}=(2\pi)^{-1}\sqrt{g/A}=5.8\,$Hz. This is a sign change in the ideal tether force, not a fitted extraction threshold.

The lumped elastic--inertial scale follows by linearizing the model during a taut interval about its gravity-stretched equilibrium. Let $Q$ denote the small absolute vertical displacement of the mass from this equilibrium and let primes denote derivatives with respect to $\tau=t\sqrt{k/m}$. The vertical equation becomes
\begin{equation}
    Q''+(\zeta+\zeta_{\rm rad})Q'+Q
    =Y_s+\zeta_{\rm rad}Y_s'.
    \label{eq:si_taut_linearization}
\end{equation}
The support-to-mass displacement transfer function is
\begin{equation}
    H(\Omega^*)=\frac{\widetilde Q}{\widetilde Y_s}
    =\frac{1+i\zeta_{\rm rad}\Omega^*}
    {1-(\Omega^*)^2+i(\zeta+\zeta_{\rm rad})\Omega^*}.
    \label{eq:si_ei_transfer}
\end{equation}
Defining the post-resonant crossover by $|H|=1$ gives
\begin{equation}
    (\Omega^*_{\rm tr})^2
    =2-\zeta^2-2\zeta\zeta_{\rm rad}.
    \label{eq:si_transmissibility_frequency}
\end{equation}
For $\zeta_{\rm rad}=0.12$ and $\zeta=8.1\times10^{-4}$--$8.1\times10^{-3}$, $\Omega^*_{\rm tr}=1.4135$--$1.4141$, essentially $\sqrt{2}$. Using Eq.~\ref{eq:si_coordinate_identity}, the corresponding no-fit guide is
\begin{equation}
    \boxed{
    \Gamma_{\rm acc,el}
    =\epsilon(\Omega^*_{\rm tr})^2\chi
    \simeq2\epsilon\chi
    =0.30\chi,}
    \label{eq:si_ei_boundary}
\end{equation}
where $\epsilon=A/L_0=0.15$. The inertia--stiffness balance occurs earlier at $\Omega^*=1$, or $\Gamma_{\rm acc}=0.15\chi$; Eq.~\ref{eq:si_ei_boundary} marks the point at which the absolute mass motion has fallen below the support motion in the linearized taut response.

The extended tangle also has a spatial transit time $L/v_{\rm eff}$. The one-cycle timing distance is $\lambda_p=v_{\rm eff}/f$, while the distances available in quarter and half cycles are $L_{\pi/2}=v_{\rm eff}/(4f)$ and $L_\pi=v_{\rm eff}/(2f)$. Using the phase delay of Eq.~\eqref{eq:si_phase_delay}, the two timing guides map to
\begin{equation}
\begin{aligned}
    f_{\pi/2}&=\frac{v_{\rm eff}}{4L},
    &\Gamma_{\rm acc,\pi/2}
       &=\frac{A}{g}\left(\frac{\pi v_{\rm eff}}{2L}\right)^2,\\
    f_\pi&=\frac{v_{\rm eff}}{2L},
    &\Gamma_{\rm acc,\pi}
       &=\frac{A}{g}\left(\frac{\pi v_{\rm eff}}{L}\right)^2.
\end{aligned}
\label{eq:si_propagation_boundary}
\end{equation}
For $v_{\rm eff}=3.2\,{\rm m\,s^{-1}}$, $L=27\,$mm and the response map's $A=0.0075\,$m, $g=9.8\,{\rm m\,s^{-2}}$, the frequencies are 29.8 and 59.5~Hz and the acceleration coordinates are 26.8 and 107.1. The shaded interval between them is the transmission crossover window.

The trajectory inference provides a separate scale check. Since $\kappa_{\rm dyn}=k/m$ has dimensions of angular frequency squared, associating its elastic time with a disturbance traversal time gives
\begin{equation}
    v_{\rm inf}\sim L\sqrt{\kappa_{\rm dyn}},
    \label{eq:si_inferred_speed}
\end{equation}
up to an order-unity factor set by distributed geometry and boundary conditions. In the representative high-frequency trial in Fig.~\ref{fig:inference_results}D, $\kappa_{\rm dyn}\simeq(0.9$--$1.1)\times10^3\,{\rm s^{-2}}$. Together with $L=20$--$30\,$mm, this gives $v_{\rm inf}\simeq0.6$--$1.0\,{\rm m\,s^{-1}}$. The inference and optical flow therefore identify the same broad $\mathcal{O}(1\,{\rm m\,s^{-1}})$ scale, although the optical-flow mean is larger by an order-unity factor. We use the direct optical-flow measurement for the quantitative propagation band.

For the 10-TW tangle, the quasi-static stiffness $k_{\rm eff}\simeq5,{\rm Nm^{-1}}$ gives $\chi\simeq510$ for $m=50~{\rm mg}$ and $L_0=5~{\rm cm}$. The corresponding natural frequency is $f_n=(2\pi)^{-1}\sqrt{k_{\rm eff}/m}=50.3~{\rm Hz}$, and the lumped transmissibility crossover occurs at $\sqrt{2}f_n=71.2~{\rm Hz}$, corresponding to $\Gamma_{\rm acc}\simeq153$. Using $v_{\rm eff}=3.2~{\rm ms^{-1}}$ and $L=27~{\rm mm}$, the quarter- and half-cycle phase delays delimit a transmission crossover window from $29.8$ to $59.5~{\rm Hz}$, corresponding to $\Gamma_{\rm acc}\simeq26.8$--$107.1$. Both bounds precede the lumped transmissibility crossover, allowing spatial dephasing to reduce coherent whole-tangle motion while the lumped model permits large center-of-mass excursions. This separation is why $k_{\rm eff}$ remains the appropriate prescribed stiffness for locating the experimental conditions on the model map, while $v_{\rm eff}$ describes a distinct spatial limit that the point-mass model cannot contain.

\section{Supplementary Videos}
\label{sec:si_videos}

\begin{description}
    \item[Supplementary Video 1.] Preparation of stochastic cotton-thread tangles by alternating clockwise and counterclockwise tumbling with a prescribed number of tumbling weights.
    \item[Supplementary Video 2.] Representative filament-extraction trial showing intermittent loading, slip and eventual detachment of the remaining tangle.
    \item[Supplementary Video 3.] High-speed disturbance-propagation recording of a 10 TW tangle at $10{,}000$ frames per second with the optical-flow field overlaid, corresponding to main-text Fig. 3B.
    \item[Supplementary Video 4.] Compacted-tangle point-mass control at representative low, intermediate and high drive frequencies.
    \item[Supplementary Video 5.] Mass-matched rigid-mesh control, retaining a soft stitched extraction thread, across the tested drive frequencies.
    \item[Supplementary Video 6.] Extended and folded stitched-mesh configurations at a fixed drive frequency, showing internal deformation in the extended mesh and center-of-mass jumps after folding.
    \item[Supplementary Video 7.] Representative tension-only model responses and the transition between continuously loaded, intermittent slack--recapture and localized high-frequency motion.
    \item[Supplementary Video 8.] Qualitative Petri-dish observation of a leech extracting one worm from an active blackworm aggregation. The leech remains posteriorly anchored while flexing and progressively wrapping around the attached worm. This submerged, nearly horizontal and biologically active configuration is presented as a natural observation motivating the local-to-global extraction question, not as a test of the gravity-mediated mechanism.
\end{description}
\bibliography{references}